\pdfoutput=1
\documentclass[aps,twocolumn,nofootinbib,superscriptaddress,floatfix]{revtex4-2}
\usepackage{amsmath,amssymb}
\usepackage{graphicx}

\newcommand{\rt}{r_{12}}
\newcommand{\dA}{d\mathcal{A}}
\newcommand{\A}{\mathcal{A}}
\newcommand{\Lie}{\mathcal{L}}
\newcommand{\eps}{\epsilon}
\newcommand{\cE}{\mathcal{E}}
\newcommand{\cB}{\mathcal{B}}

\begin{document}

\title{Horizon entropy does not by itself determine\\a force between black holes}

\author{Zhi-Wei Wang}
\email{zhiweiwang.phy@gmail.com}
\affiliation{College of Physics, Jilin University, Changchun, 130012,
People's Republic of China}
\affiliation{Department of Computer Science, University of York,
York YO10 5GH, United Kingdom}
\author{Samuel L. Braunstein}
\email{sam.braunstein@york.ac.uk}
\affiliation{Department of Computer Science, University of York,
York YO10 5GH, United Kingdom}

\date{\today}

\begin{abstract}
The Bekenstein--Hawking law fixes the total entropy of a horizon. Membrane,
holographic, and entropic-force pictures go further and follow the entropy of
individual horizon elements as a configuration changes. We show that this step
needs structure that the horizon geometry does not supply. Along a family of
horizons labeled by $\lambda$, a relabeling of horizon points by a tangential
field $v^A$ shifts the temperature-weighted entropy variation
$\int T\,\partial_\lambda(dS)$ exactly by $-\int v^AD_AT\,dS$, so the horizon
data determine it only when the temperature is uniform. We test this on two
black holes in exact Brill--Lindquist data. At fixed physical masses each
horizon area is independent of the separation through five relative orders; the
low-order equilibrium force reported earlier comes from holding the bare
parameters fixed instead. Weighting the redistribution of area by the
nonuniform Noether-charge temperature gives a candidate repulsion equal to
$(\nu/8)(GM/r_{12})^3$ times the Newtonian attraction, for total mass $M$,
symmetric mass ratio $\nu$, and coordinate separation $r_{12}$. As a static
interaction it would shift the third post-Newtonian coefficient of the
circular-orbit binding energy by $5\nu/48$, in conflict with general
relativity; the static transverse-traceless part of the physical slice leaves
this unchanged at that order. The identity resolves the conflict. Through
fourth post-Newtonian order the candidate lies entirely in the dipole of the
area distribution, which the uniform part of the companion's field generates
as a M\"obius map of the labels, and which vanishes in the area-balanced
conformal frame, where the functional begins at fifth order with a coefficient
that a further relabeling can still change. The marginally trapped tube of the data
and the tidally deformed horizon of a post-Newtonian binary, both of which do
supply an identification, transport no dipole at the order the candidate
requires. When the temperature is uniform the construction returns a first-law
term: for the Israel--Khan pair it is conjugate to the strut tension along a
path of fixed total mass and vanishes along a path of fixed horizon areas.
\end{abstract}

\maketitle

\section{Introduction}
\label{sec:intro}

The Bekenstein--Hawking law~\cite{Bekenstein1973,Hawking1975,Bardeen1973}
\begin{equation}
S=\frac{\A}{4G}
\label{eq:BH}
\end{equation}
assigns an entropy to a horizon as a whole. We use units with
$c=\hbar=k_B=1$ and retain $G$. Gravitational-wave observations now test
Hawking's area theorem for that total: in the event GW250114 the horizon area
of the remnant exceeds the sum of the areas of the two holes that merged, with
high credibility~\cite{GW250114}, following the earlier test with
GW150914~\cite{Isi2021}. These measurements concern total horizon areas
inferred from the signal, that is, geometry. They do not probe how entropy is
distributed over a horizon, or how it is carried from one configuration to
another.

Many uses of the law are nonetheless local. The membrane paradigm attaches
surface densities to the horizon~\cite{Thorne1986}; dynamical-horizon and
entropy-current constructions distribute entropy over horizon
cross-sections~\cite{AshtekarKrishnan2004,HollandsWaldZhang2024,EntropyCurrent2021,Reparam2023,Kar2024};
the expectation that information falling into a hole is recorded on the horizon
where it fell in is local by construction; and entropic-force arguments
differentiate an entropy as a configuration
changes~\cite{Verlinde2011,Morozov2010,vanPutten2012}. The last use rests on two
assumptions beyond Eq.~(\ref{eq:BH}). The first is that an element of horizon
area carries an element of entropy. The second, rarely stated, is that the same
element can be followed from one configuration to the next, so that its entropy
can be transported. Entropy-current constructions make the first precise; there
the density and the spatial current each change under a reparametrization of
the generators, while the net production does not~\cite{Reparam2023,Kar2024}.
Here we ask whether the horizon geometry supplies the second.

The natural probe is mechanical, and it follows from the two assumptions rather
than from any view of what gravity is. In the principle of virtual work an
entropic generalized force is a one-form on the space of admissible variations,
$\delta W_S=T\,dS(\delta q)$. For an isothermal system this is the entropy part
of the generalized force $Q_q=-(\partial F_{\rm Helm}/\partial q)_T$, and when
the energetic part does not depend on $q$ it is all of it,
$Q_q=T(\partial S/\partial q)_T$. A stretched polymer is the familiar example,
in which the derivative is taken on the physical configuration space of the
chain~\cite{deGennes1979,Doi2013}. A local horizon version must specify not
only an entropy density on each horizon but also the variation with respect to
which that density is differentiated, that is, which point of one horizon is
the same as a given point of the next.

Two black holes provide a concrete test. In exact Brill--Lindquist data,
weighting the change of each horizon element's entropy by the local
Noether-charge temperature gives a candidate force between the holes, a
repulsion that enters at third post-Newtonian (3PN) order. Were it an
additional static interaction, it would shift the 3PN coefficient of the
gauge-invariant circular-orbit binding energy by $\delta e_3=5\nu/48$, with
$\nu$ the symmetric mass ratio, and the 3PN conservative dynamics derived from
Einstein's equations leaves no room for such a
term~\cite{DJS2001,BlanchetFaye2001,DJS2001dimreg,Blanchet2014}. The
construction must therefore fail somewhere.

It fails at the second assumption. An exact identity shows that the
temperature-weighted entropy variation changes under a relabeling of horizon
points by an amount set by the temperature gradient, so the horizon data fix it
only when the temperature is uniform. In Brill--Lindquist data, the 3PN and 4PN
terms of the candidate lie entirely in the dipole of the horizon's area
distribution. That dipole is not horizon geometry: it is produced by the
uniform part of the companion's field, which acts on the horizon as a
M\"obius map of the labels, and the conformal frame in which the area
distribution is balanced, unique up to rotations, removes it. Two constructions in which the dynamics does supply an
identification, the marginally trapped tube of the data and the tidally
perturbed horizon of a post-Newtonian binary, transport no dipole at the
required order. When the temperature is uniform, as for static holes held apart
by a strut, the construction is well defined, and it returns a first-law term
whose meaning depends on the thermodynamic path rather than a force between the
holes.

Forces of this kind between black holes have been
reported~\cite{Morozov2010,vanPutten2012}, and Verlinde has proposed that
gravitation is itself entropic in character~\cite{Verlinde2011}. Neither is our
starting point. The relabeling identity and an account of its binary
realization appear in a companion paper~\cite{Companion}; the present paper gives the full derivations, the
thermodynamic-path analysis, the dynamical checks, and the post-Newtonian
comparison.

Section~\ref{sec:relabel} derives the identity in general.
Section~\ref{sec:setup} fixes the data, the horizon, and the temperature.
Section~\ref{sec:rigid} shows that the horizon area at fixed physical mass is
rigid through five relative orders, and Sec.~\ref{sec:smarr} establishes an
exact Smarr relation for the interacting pair. Section~\ref{sec:candidate}
constructs the candidate and shows that it depends on the identification.
Section~\ref{sec:corroboration} records what the candidate would commit one to
if it were an interaction, including the effect of the static
transverse-traceless part of the physical slice. Section~\ref{sec:isometry}
shows that its 3PN and 4PN terms are a frame mode, and
Sec.~\ref{sec:dynamics} compares it with two dynamical transport laws.
Section~\ref{sec:uniform} treats the uniform-temperature case through the first
law, and Sec.~\ref{sec:discussion} discusses what has and has not been shown.
The appendices collect the derivations and the independent checks.

\section{Relabeling and local work}
\label{sec:relabel}

Let $H_\lambda$ be a one-parameter family of closed horizon cross-sections, for
instance the horizons of a sequence of configurations, each carrying an entropy
two-form $dS_\lambda$ and a temperature field $T_\lambda$. Differentiating
$dS_\lambda$ with respect to $\lambda$ requires a rule that says which point of
$H_{\lambda+d\lambda}$ is the same as a given point of $H_\lambda$; we call such
a rule an identification. Concretely, an identification is a smooth family of
labelings of $H_\lambda$ by a fixed sphere, and $\partial_\lambda$ is taken at
fixed label. Given one, the natural local version of $T\,dS/d\lambda$ is
\begin{equation}
\mathcal W_\lambda=\int_{H_\lambda}T_\lambda\,\partial_\lambda(dS_\lambda).
\label{eq:W}
\end{equation}

Changing the identification composes each labeling with a diffeomorphism of the
sphere that depends on $\lambda$. Its rate of change is a tangential field
$v^A$ on $H_\lambda$, the label velocity, and the new derivative at fixed label
is $\partial_\lambda+\Lie_v$ on cross-section fields. Since $dS$ is a top form,
Cartan's formula gives $\Lie_v dS=d(i_vdS)$, and since
$dT\wedge i_vdS=(v^AD_AT)\,dS$ on a two-surface, with $D_A$ the covariant
derivative of the cross-section, Stokes' theorem on the closed surface gives
\begin{equation}
\mathcal W'-\mathcal W=\int T\,\Lie_v(dS)=-\int v^AD_AT\,dS .
\label{eq:relabel}
\end{equation}
Equation~(\ref{eq:relabel}) is exact. It uses no field equation and holds for
any density and any temperature field. Appendix~\ref{app:relabel} gives the
same identity in the axisymmetric form used below, together with a closed-form
example.

The integrated charge $Q_\lambda=\int T_\lambda\,dS_\lambda$ is unaffected,
because in
\begin{equation}
\partial_\lambda Q_\lambda
=\int T\,\partial_\lambda(dS)+\int(\partial_\lambda T)\,dS
\label{eq:Qsplitmain2}
\end{equation}
the second term shifts by $+\int v^AD_AT\,dS$ under the same relabeling, and the
two shifts cancel. What changes is the stand-alone transport
term~(\ref{eq:W}). It is invariant under all smooth relabelings if and only if
$D_AT=0$: if $D_AT$ is nonzero on an open set, the choice $v^A=-c\,\varphi\,D^AT$
with $\varphi\ge0$ a bump function supported there shifts $\mathcal W$ by
$c\int\varphi\,|DT|^2dS$, which takes any real value as $c$ varies.

Two readings follow. If the physics supplies an identification, such as a
material velocity, a membrane flow, or the generators of an evolving horizon,
then Eq.~(\ref{eq:W}) can be meaningful at nonuniform $T$, and
Eq.~(\ref{eq:relabel}) says how its value depends on that choice. If nothing
supplies one, Eq.~(\ref{eq:W}) is fixed by the family
$(H_\lambda,dS_\lambda,T_\lambda)$ only at uniform temperature, which is what
the zeroth law delivers in equilibrium. Away from equilibrium a local entropy
can then be assigned but not transported without additional structure.

There is a second freedom, in the density itself. Equation~(\ref{eq:BH}) fixes
only a total, so $dS\to dS+d\omega$ with $\oint d\omega=0$ returns the same
entropy, and on a closed two-surface a top form of vanishing integral is exact.
Whether such a form is an admissible Jacobson--Kang--Myers ambiguity of the
Noether charge is a separate question~\cite{Wald1993,IyerWald1994,JacobsonKangMyers1994},
since the admissible terms there are constrained by locality and covariance.
Recent constructions of dynamical black-hole entropy do not fix the density in
the regime studied here: they dispose of the freedom by integrating it away,
they hold at first order in perturbations of a stationary black hole while the
effect below is second order, and the one nonperturbative treatment excludes
time-symmetric data by hypothesis (Appendix~\ref{app:density}). We hold the
density at $\dA/4G$ throughout. Equation~(\ref{eq:relabel}) holds whatever the
density, and nothing below depends on the second freedom.

\section{Data, horizons and temperature}
\label{sec:setup}

\subsection{Setting}

We make one choice of setting and record one property of the data.

The choice is that the thermodynamic surface of a black hole is its marginally
trapped surface rather than its event horizon. This is the quasi-local view now
broadly adopted~\cite{AshtekarKrishnan2004,Booth2005,GourgoulhonJaramillo2006},
and it is the setting in which the local temperature used below was
constructed. Our calculation compares configurations rather than following a
history, which is also the character of the observational test, since GW250114
compares inferred initial and final total areas.

The property is that the horizons are instantaneously nonexpanding. The
spacetime is vacuum, so no matter crosses the horizon, but the shear supplies a
gravitational flux, which is what tidal heating is, and we return to it in
Sec.~\ref{sec:dynamics}. What holds at a moment of time symmetry is that the
marginally trapped surface has vanishing expansion, so its instantaneous area
derivative vanishes exactly, while its shear need not vanish. Instantaneously
nonexpanding is not the same as isolated, and the difference is essential. On an
isolated horizon the expansion $\theta_{(k)}$ of the outgoing null normal $k$
(defined below) vanishes on an interval, so in vacuum the Raychaudhuri equation
reduces to $|\sigma^{(k)}|^2=0$, with $\sigma^{(k)}$ the shear of $k$, and the
further structure of a weakly isolated horizon yields the zeroth law and a
uniform surface gravity~\cite{AshtekarKrishnan2004}. Our horizons are
marginally trapped at one instant with $\sigma^{(k)}\neq0$. They are therefore
not isolated, the zeroth law does not apply, and the surface gravity is not
uniform. That is not a defect of the construction; it is the feature on which
the calculation turns.

\subsection{Brill--Lindquist data}

We work with Brill--Lindquist data for two nonrotating holes at a moment of
time symmetry~\cite{Brill1963,Misner1963,Cook2000}. The extrinsic curvature
vanishes and the three-metric is conformally flat,
\begin{equation}
\gamma_{ij}=\psi^4\delta_{ij},\qquad
\psi=1+\frac{G\mu_1}{2\rho_1}+\frac{G\mu_2}{2\rho_2},
\label{eq:psi}
\end{equation}
with $\rho_i$ the flat distance from the $i$th puncture and $\rt$ the flat
coordinate separation of the punctures. The data are exact, and every power of
$G$ below is an expansion of an exact object rather than of the field
equations.

The parameters $\mu_i$ are not the physical masses. Inverting about the $i$th
puncture by a Kelvin transform exhibits a second asymptotically flat region, and
the mass measured there is
\begin{equation}
m_i=\mu_i\Bigl(1+\frac{G\mu_j}{2\rt}\Bigr),\qquad j\neq i,
\label{eq:misner}
\end{equation}
exactly~\cite{Misner1963}, while the total energy is $E_{\rm ADM}=\mu_1+\mu_2$,
also exactly. The $\mu_i$ are thus bare parameters from which the interaction
energy has already been deducted, and $m_i$ is what an observer in the $i$th
region calls the mass of that hole. Which of the two is held fixed as the
separation varies matters for everything that follows. We write
$M=m_1+m_2$, $\nu=m_1m_2/M^2$ and $\eps\equiv GM/\rt$.

Let $\hat{u}$ be the unit normal to the slice and $\hat{N}$ the outward unit
normal to a closed two-surface within it, and define the null normals
$k=\hat{u}+\hat{N}$ and $n=\hat{u}-\hat{N}$, so that $k\cdot n=-2$. Their
expansions are $\theta_{(k)}=q^{ij}(D_i\hat{N}_j-K_{ij})$ and
$\theta_{(n)}=-q^{ij}(D_i\hat{N}_j+K_{ij})$, where $q_{ij}$ is the induced metric,
$D_i$ the covariant derivative of the slice, and $K_{ij}$ its extrinsic
curvature. At time symmetry $K_{ij}=0$, so $\theta_{(k)}=-\theta_{(n)}$ is the
mean curvature of the surface, and the marginally outer trapped surface (MOTS)
is the outermost minimal surface around each puncture. We write it
$\rho_i=\varrho_i(\theta)$, with $\theta$ the polar angle about the puncture
measured from the direction of the companion; the data are axisymmetric about
the line of centers.

\subsection{Temperature and lapse}

A temperature must be assigned pointwise, and the data admit no Killing field.
We use the Noether-charge construction of
Refs.~\cite{WangBraunstein2018,WangBraunstein2026}, in which
\begin{equation}
\kappa=\xi_{a;b}\hat{N}^a\hat{u}^b,\qquad T=\frac{\kappa}{2\pi},
\label{eq:Tlocal}
\end{equation}
with $\xi^a=\mathcal N\hat{u}^a$ the evolution vector of the Eulerian
observers. On the slice this is $\kappa=\hat{N}^iD_i\mathcal N$. That $\kappa$ is
a scalar field on the horizon, defined pointwise and requiring no Killing
symmetry, is established in Ref.~\cite{WangBraunstein2026}.

This yields a definite field only once the evolution vector, and hence the lapse
$\mathcal N$, has been specified, and it is not an observer-independent
thermodynamic temperature; Sec.~\ref{sec:isometry} shows that its anisotropy is
a property of the Eulerian observers, momentarily at rest with the holes. It also differs from the
proper temperature of a stretched horizon. At a fixed small proper distance
$\rho_*$ from a stationary horizon a local Rindler observer sees
$T_{\rm prop}\simeq(2\pi\rho_*)^{-1}$, approximately uniform over the
horizon~\cite{Unruh1976,Thorne1986}, and in the stationary local-Rindler limit
the Tolman redshift gives $\mathcal NT_{\rm prop}=\kappa/2\pi$~\cite{Tolman1930}.
The $T$ of Eq.~(\ref{eq:Tlocal}) is therefore the redshifted weight associated
with the chosen evolution, normalized at infinity. It is the weight that the
construction under test uses, and it is fixed as follows.

Maximal slicing requires $\mathcal N=\chi/\psi$ with $\chi$ harmonic in the flat
metric. Harmonic functions regular at infinity may carry multipoles of any order
about each puncture, and these must be excluded, since a term
$c\cos\theta/\rho_1^2$ is harmonic and contributes a freely tunable
$\cos\theta$ term to $\kappa$, which the Smarr relation of
Sec.~\ref{sec:smarr}, a monopole integral, cannot detect. What excludes it is
the second asymptotic end:
under the Kelvin transform about the first puncture that term grows linearly, so
the lapse would diverge there. Requiring $\mathcal N\to1$ at infinity and
$\mathcal N\to-1$ at each inner end, the odd normalization of puncture slicings,
therefore fixes $\chi$ completely and gives (Appendix~\ref{app:lapse})
\begin{equation}
\chi=2-\psi,\qquad \mathcal N=\frac{2}{\psi}-1 .
\label{eq:chi}
\end{equation}
The sign at the inner ends matters: $\mathcal N\to+1$ there would instead give
$\chi=\psi$, $\mathcal N\equiv1$ and $\kappa\equiv0$. For a single hole
Eq.~(\ref{eq:chi}) gives $\mathcal N=0$ on the horizon and $\kappa=1/4Gm$, the
Schwarzschild surface gravity for a Killing field normalized at infinity and
the value required by the isolated-horizon first law
$\delta E=\kappa\,\delta\A/8\pi G$. The zeroth law fails for the binary, for
the reason given above, but the first-law normalization does not.

Expanding about the first horizon,
\begin{equation}
\kappa_1=\frac{1}{4Gm_1}-\frac{m_2}{8m_1\rt}
-\frac{3Gm_2}{16\rt^2}\cos\theta+\cdots ,
\label{eq:kappaexp}
\end{equation}
so the companion lowers this temperature weight on the near side and raises it
on the far side. Appendix~\ref{app:setup} derives both terms. The constant is
the companion's potential, which redshifts the hole as seen from infinity. The
dipole is the companion's uniform Newtonian field at the hole, seen through a
lapse normalized at infinity; it is proportional to the acceleration the hole is
about to acquire, as seen in the frame in which it is momentarily at rest.

\begin{figure}[tbp]
\includegraphics[width=\columnwidth]{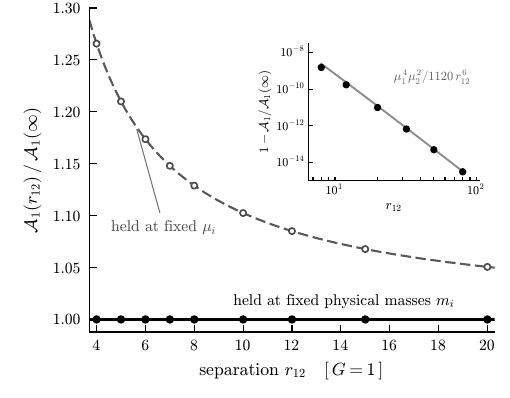}
\caption{Rigidity of the horizon area, computed from the exact minimal-surface
solution of the Brill--Lindquist data rather than from any expansion. Held at
fixed conformal parameters $\mu_i$ (open circles) the area of one hole appears
to grow as the companion approaches; held at fixed physical masses $m_i$
(filled circles) it is constant on the scale of the main panel through five
relative orders. The apparent growth is the factor
$(1+G\mu_j/2\rt)^2=m_i^2/\mu_i^2$ of Eq.~(\ref{eq:misner}) (dashed curve) and
carries no geometry. Inset: the residual at fixed physical masses (points)
against the leading term of Eq.~(\ref{eq:area}) written in the bare
parameters (line), which the points approach as the separation grows. Open
circles are computed with $\mu_1=\mu_2=1$ and filled circles with
$m_1=m_2=1$; lengths are in units with $G=1$ and the corresponding unit mass.
All points are unchanged, to the precision plotted, between Legendre
truncations 16 and 28.}
\label{fig:area}
\end{figure}

\section{The horizon area is rigid}
\label{sec:rigid}

Solving the minimal-surface condition about the first puncture and integrating
the induced area gives (Appendix~\ref{app:area})
\begin{equation}
\A_i=16\pi G^2m_i^2
\Bigl(1-\frac{G^6m_i^4m_j^2}{1120\,\rt^{6}}+\cdots\Bigr),
\qquad j\neq i ,
\label{eq:area}
\end{equation}
where inside the correction $\mu$ and $m$ may be interchanged, the difference
entering only at relative order $G^7$. The separation dependence has canceled
through five relative orders. It has not canceled altogether: the residue is a
small decrease of the area at relative order $G^6$, to which we return in
Secs.~\ref{sec:corroboration} and~\ref{sec:isometry}. What the calculation
establishes is that the entire response of the total area through five relative
orders is algebraic and removed by Eq.~(\ref{eq:misner}), and that this holds
for each hole separately rather than only for the sum. The intrinsic geometry is
not unchanged: the tidal quadrupole distorts it at relative order $G^3$, as
Sec.~\ref{sec:isometry} makes explicit.

Two things make the cancellation more than a low-order accident. First, the
change of mass variables is exact. Subtracting the two relations of
Eq.~(\ref{eq:misner}) gives $m_1-m_2=\mu_1-\mu_2$, so
$\mu_i=m_i-\tfrac12E_b$, where $E_b\equiv M-E_{\rm ADM}=G\mu_1\mu_2/\rt$ is the
binding energy. It satisfies a quadratic, whence
\begin{equation}
E_b=\frac{GM+2\rt-\sqrt{(GM+2\rt)^2-4G^2m_1m_2}}{G},
\label{eq:Pexact}
\end{equation}
which is $Gm_1m_2/\rt$ at leading order.
Holding the $m_i$ fixed is therefore a well-defined path through the family,
along which the prefactor $16\pi G^2\mu_i^2(1+G\mu_j/2\rt)^2$ of the area
written in the bare parameters is $16\pi G^2m_i^2$ at every order.

Second, the order at which the geometry does contribute can be seen without
computing it. Expanded about the first puncture, the companion's field supplies
a constant, a term linear in the Cartesian coordinate $z$ along the line of
centers, and a quadrupole, at successive orders in $1/\rt$. The constant is
absorbed by Eq.~(\ref{eq:misner}). The linear term is the uniform field, and
Sec.~\ref{sec:isometry} shows that the constant and linear terms together are
exactly the image of the isolated throat under a finite special conformal map,
so that they change no invariant at any order and the whole remainder of the
companion's field is a regular harmonic function with only $\ell\ge2$
components about the puncture. The quadrupole is suppressed twice over. Its
direct contribution to the area integrates to zero because it carries no
monopole, and the displacement of the surface that it induces changes the area
only at second order because the area is stationary on a minimal surface. What
survives is quadratic in a quadrupole of relative amplitude $O(G^3)$, and so of
relative order $G^6$, which is what Eq.~(\ref{eq:area}) exhibits.

The coefficient has been obtained in three ways, the last independent of the
series: from the $O(G^6)$ term of the force series of Appendix~\ref{app:force};
from direct numerical evaluation of the area using the expansions of
Appendix~\ref{app:area}; and from a variational solution for the minimal
surface that uses neither (Appendices~\ref{app:area} and~\ref{app:rebuild}).

The consequence is that $\partial S_i/\partial\rt=O(G^7)$. The equilibrium
entropic force $T\,\partial S/\partial\rt$ therefore enters at $O(G^6)$, that is
at fifth post-Newtonian order, two powers of $G$ above the candidate of
Sec.~\ref{sec:candidate}, and any entropic effect at lower order must come from
the redistribution of area rather than from its total.

This also locates the earlier reports. Van Putten~\cite{vanPutten2012} computes
the same area in the same data, and our expansion reproduces his term by term
when written in the $\mu_i$ (Appendix~\ref{app:area}). What we add is that all
of its separation dependence through relative order $G^5$ is the factor
$m_i^2/\mu_i^2$, which by Eq.~(\ref{eq:misner}) is algebraic and carries no
geometry (Fig.~\ref{fig:area}). The construction of Morozov~\cite{Morozov2010}
is of a different kind, and our result does not bear on it: he defines the
horizon by a Newtonian escape-velocity condition rather than as a marginally
trapped surface, and notes himself that his argument is largely dimensional.

\section{An exact Smarr relation}
\label{sec:smarr}

Define $Q[\Sigma]=(8\pi G)^{-1}\oint_\Sigma\kappa\,\dA$, which in the units of
Eq.~(\ref{eq:BH}) is $\oint_\Sigma T\,dS$. By Eq.~(\ref{eq:chi}),
$\psi^2\partial_i\mathcal N=-2\partial_i\psi$ pointwise, and combining this with
$\hat{N}^i=\psi^{-2}\hat{N}^i_{\rm f}$ and $\dA=\psi^4dA_{\rm f}$, where the
subscript denotes the flat metric, every conformal factor cancels:
\begin{equation}
\kappa\,\dA=-2\,\partial_{\hat{N}_{\rm f}}\psi\;dA_{\rm f} .
\label{eq:flatflux}
\end{equation}
The curved-space integral is a flat-space flux. Since $\psi$ is harmonic away
from the punctures, the flux through any surface enclosing only the $i$th
puncture may be evaluated on a small sphere about it, where only the term
$G\mu_i/2\rho_i$ contributes, giving $\oint\kappa\,\dA=4\pi G\mu_i$. Hence
\begin{equation}
Q_i=\frac{\mu_i}{2}\quad\text{exactly},\qquad 2\sum_iQ_i=E_{\rm ADM},
\label{eq:Qexact}
\end{equation}
for the interacting pair at any separation, with no expansion.

The factor of two is the ordinary Komar factor. For a single Schwarzschild hole
$Q=\kappa\A/8\pi G=m/2$, that is $m=2TS$, the standard Smarr relation, and
Eq.~(\ref{eq:Qexact}) says that the interacting pair obeys the same relation
exactly. No inference about a mechanical variation follows from it, however,
since $d(TS)=T\,dS+S\,dT$ and the Smarr relation constrains only the product.
The supported static example of Sec.~\ref{sec:uniform} illustrates this: there
$T_iS_i$ is constant along the path considered while $T_i\,dS_i$ is not,
the difference being carried by $S_i\,dT_i$. What the construction returns is
fixed by the first law rather than by Smarr.

\begin{figure}[tbp]
\includegraphics[width=\columnwidth]{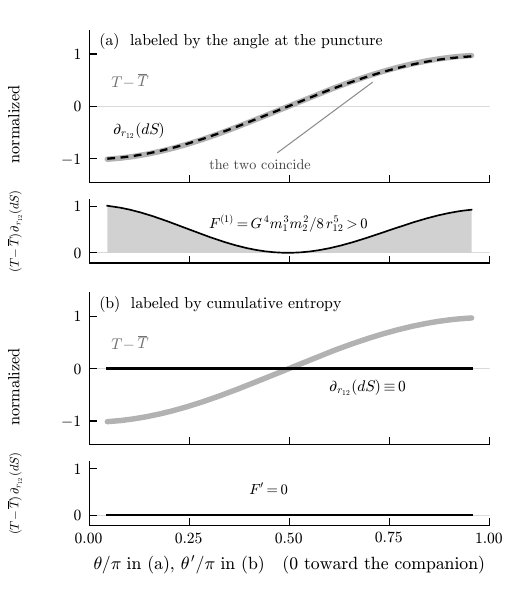}
\caption{The candidate is a product of two dipoles. Curves are extracted from
the exact minimal-surface solution at $\rt=30$ for $G=m_1=m_2=1$ and normalized
to unit amplitude; the temperature anisotropy $T-\bar{T}$, with $\bar{T}=Q/S$ the
entropy-weighted mean, is itself of relative order $(Gm/\rt)^2$. (a) Labeled by the angle at the
puncture, the entropy-density derivative has the same angular dependence as the
temperature dipole, so their product, shown in the lower strip per unit
$\cos\theta$ for the first horizon alone, is positive throughout; its integral
$F^{(1)}=G^4m_1^3m_2^2/8\rt^5$ is that horizon's contribution to
Eq.~(\ref{eq:Fpair}). (b) Labeled by cumulative entropy, the density derivative
vanishes, and so does the product, the horizontal coordinate now being the
angle $\theta'$ for which $(1-\cos\theta')/2$ is the fraction of the entropy
enclosed. The two lower strips share one scale. The curves are the same at
Legendre truncations 16 and 28.}
\label{fig:dipoles}
\end{figure}

\section{The candidate force}
\label{sec:candidate}

\subsection{Definition and value}

The total area does not change but its distribution does, and by
Eq.~(\ref{eq:kappaexp}) the temperature is not uniform. Weighting the one by the
other gives the natural candidate. Take $\lambda=\rt$ in Eq.~(\ref{eq:W}),
identify the points of each horizon by their puncture angles $(\theta,\phi)$,
hold the physical masses fixed, and sum over the two horizons:
\begin{equation}
F\equiv\sum_i\int T_i\,\partial_{\rt}\Bigl(\frac{\dA_i}{4G}\Bigr)
\Big|_{\theta,\phi,m_1,m_2}.
\label{eq:F}
\end{equation}
A positive $F$ means that the temperature-weighted entropy grows with
separation, so that the candidate pushes the holes apart. Evaluating
Eq.~(\ref{eq:F}) on the data of Sec.~\ref{sec:setup} gives
(Appendix~\ref{app:force})
\begin{equation}
F=\frac{G^4m_1^2m_2^2(m_1+m_2)}{8\,\rt^5}+O(G^5),
\label{eq:Fpair}
\end{equation}
of which the first horizon contributes $F^{(1)}=(m_1/M)F$, or
\begin{equation}
\frac{F}{F_{\rm N}}=\frac{\nu}{8}\Bigl(\frac{GM}{\rt}\Bigr)^{3}=\frac{\nu}{8}\,\eps^3,
\qquad F_{\rm N}=\frac{Gm_1m_2}{\rt^2},
\label{eq:ratio}
\end{equation}
which is third post-Newtonian order: $3\times10^{-5}$ of the Newtonian
attraction at $\rt=10\,GM$ for equal masses. The candidate depends on the
separation alone, so it derives from a potential (Sec.~\ref{sec:corroboration}),
and it carries no $\hbar$: restoring units, $T\propto\hbar$ and
$S\propto\hbar^{-1}$, so that $T\,dS=\kappa\,\dA/8\pi G$ is classical.

Holding the physical masses fixed is essential. Holding the bare parameters
$\mu_i$ fixed instead follows a path of constant $E_{\rm ADM}$ and produces the
Newtonian-scale term reported in Ref.~\cite{vanPutten2012}. That term
re-expresses the ordinary interaction energy through a path-dependent entropy
derivative and is not an additional horizon force (Sec.~\ref{sec:path} and
Appendix~\ref{app:rebuild}). The fixed-mass derivative isolates the effect of
redistributing the area.

\subsection{Dependence on the identification}

Equation~(\ref{eq:F}) differentiates an entropy density with respect to the
separation. It therefore compares two distinct configurations, and evaluates the
density at the same point in both. Which point of the horizon at one separation
counts as the same point at another is not a side question but the operation
the formula performs, and nothing in the definition supplies the
correspondence. The two horizons are surfaces in different curved
three-geometries, and the Brill--Lindquist family is a family of distinct
spacetimes rather than a history. Nor is this a choice of coordinates in the
usual sense: relabeling a single horizon carries no content, and what is at
issue is how the labeling changes from one member of the family to the next.
The choice of thermodynamic surface made in Sec.~\ref{sec:setup} does not help,
since it is a statement about each configuration separately.

By axisymmetry we may integrate over $\phi$ and write
$dS=\tilde{s}(\theta,\rt)\,d\theta$. For any smooth relabeling
$\theta=g(\theta',\rt)$ that fixes the poles, with label velocity
$v=\partial_{\rt}g$, Eq.~(\ref{eq:relabel}) becomes
(Appendix~\ref{app:relabel})
\begin{equation}
F'=F-\int(\partial_\theta T)\,v\,dS ,
\label{eq:relabelaxi}
\end{equation}
with the integral summed over the two horizons, each with its own label
velocity. Since $\partial_\theta T\neq0$ by Eq.~(\ref{eq:kappaexp}), a suitable $v$ gives
$F'$ any value. One choice is natural. Labeling the points of each horizon by
the entropy enclosed between the pole and the point makes the density
separation independent wherever the total is, so that $F'=0$ identically
[Fig.~\ref{fig:dipoles}(b)]. Such a labeling exists with no symmetry assumption:
by Moser's theorem in its family version~\cite{Moser1965}, a smooth family
$dS_\lambda$ of area forms of fixed total on a sphere admits a smooth isotopy
$g_\lambda$ with $g_\lambda^*dS_\lambda=dS_0$, and by
Eq.~(\ref{eq:area}) the totals are fixed through relative order $G^5$.

What remains invariant is the integrated charge. By Eq.~(\ref{eq:Qexact}) its
variation at fixed physical masses is fixed exactly,
$dQ_i/d\rt=-\tfrac14\,dE_b/d\rt$, the same for both holes, so that
$\sum_idQ_i/d\rt=\tfrac12\,dE_{\rm ADM}/d\rt$. In the split of
Eq.~(\ref{eq:Qsplitmain2}) this invariant variation and the
identification-dependent term $\int(\partial_{\rt}T)\,dS$ agree term by term
through $O(G^3)$ and first differ at $O(G^4)$, where their difference is the
candidate (Appendix~\ref{app:force}). The obstruction is therefore to assigning a
stand-alone meaning to the transport term, not to the total variation of the
charge.

Equation~(\ref{eq:relabelaxi}) shows that the candidate is under-determined. It
does not by itself show that the candidate is spurious, since the puncture
labeling is the one in which the data are written, and one might hope that it,
or some other labeling, is preferred. Section~\ref{sec:isometry} shows that at
the orders where the candidate appears there is a labeling fixed by the
intrinsic geometry, and that it gives zero. First we record what the
puncture-label value would commit one to.

\section{What the puncture labeling would commit one to}
\label{sec:corroboration}

Suppose the puncture-label value were taken as a physical interaction. It
depends only on the two masses and the separation, so it would be an additional
universal conservative interaction between the holes. The conservative
two-body dynamics of general relativity is already fixed at the order in
question, having been derived independently in two gauges and shown to
agree~\cite{DJS2001,BlanchetFaye2001,Blanchet2014}, and because internal
structure is effaced from that dynamics~\cite{Damour1987}, an interaction that
depends only on the masses and the separation would have to appear in it.

The candidate is static, with potential $V=G^4m_1^2m_2^2M/32\,\rt^4$. In the
reduced variables of Ref.~\cite{DJS2000inv}, separations in units of $GM$,
written $\hat{r}$, and energies in units of the reduced mass $m_1m_2/M$, this is
$\nu/32\hat{r}^4$, which has the form of the third post-Newtonian static term
$\nu\,\omega_{\rm static}/\hat{r}^4$ of the
Hamiltonian~\cite{JS1998,JS1999,DJS2000inv}, with
\begin{equation}
\delta\omega_{\rm static}=\tfrac{1}{32}.
\label{eq:omegas}
\end{equation}
That parameter was the last undetermined number in the 3PN dynamics. It is
normalized so that $\omega_{\rm static}=0$ is the static sector with vanishing
momenta and vanishing transverse-traceless part, which is to say
time-symmetric conformally flat data~\cite{DJS2000inv}, and dimensional
regularization gives zero for it in general relativity~\cite{DJS2001dimreg}.

Two points about the identification of Eq.~(\ref{eq:omegas}) need care. The
first concerns the radial variable. Our $\rt$ is the Brill--Lindquist coordinate
separation, while the Hamiltonian uses the separation of the ADM transverse
gauge. The two agree at leading order, and a change of the radial variable at
relative order $GM/\rt$ alters the coefficient of a term that is already of 3PN
order only at 4PN, so Eq.~(\ref{eq:omegas}) is unambiguous at the order at which
it is used. What is not invariant is the coefficient of an isolated static
$1/\hat{r}^4$ monomial within the 3PN Hamiltonian itself: the canonical
transformations available at that order move it, which is why the
Brill--Lindquist and Misner--Lindquist normalizations differ and why the
reference had to be defined on the former. The ambiguity parameters enter the
invariant dynamics only through
$\sigma(\nu)\equiv\nu\,\omega_{\rm static}+\nu^2\omega_{\rm kinetic}$, where
$\omega_{\rm kinetic}$ is the coefficient of the kinetic ambiguity term of the
same order, and Appendix~\ref{app:pn} gives a short canonical-transformation
derivation of this unit relative coefficient~\cite{DJS2000inv,DJS2000poincare}.
Time-symmetric data determine only the static coefficient. Treated as a static
interaction, the candidate has $\delta\omega_{\rm static}=1/32$ and
$\delta\omega_{\rm kinetic}=0$, hence $\delta\sigma=\nu/32$, and the invariant
energy of circular orbits changes by
\begin{equation}
\delta e_3=\tfrac{10}{3}\,\delta\sigma=\frac{5\nu}{48}.
\label{eq:de3}
\end{equation}
A kinetic counterpart could change this value but could not remove it, since a
pure number $\delta\omega_{\rm kinetic}$ cannot cancel $\nu/32$ in
$\delta\sigma$ for every $\nu$. The shift therefore cannot be removed by a
canonical transformation.

The second point concerns the slice. The Brill--Lindquist slice is not a slice
of the post-Newtonian spacetime, since the physical vanishing-momentum slice
carries a static transverse-traceless part at relative second post-Newtonian
order. Appendix~\ref{app:tt} shows that this part changes the candidate first
at fourth post-Newtonian order, and at that order only through relabelings
(Sec.~\ref{sec:isometry}). The candidate on the physical slice therefore equals
its Brill--Lindquist value through third post-Newtonian order, and the two
configurations agree in the sense of the reference normalization without being
identical.

There is accordingly no room for such a term in the established third
post-Newtonian conservative dynamics of general relativity. This is a
consistency requirement within general relativity, not an observational bound;
the premise it rests on is that the conservative two-body dynamics at that
order is complete. It says nothing about theories other than general
relativity, in which an additional interaction of this form would not be
excluded by the same argument. Within general relativity the conclusion is that
the puncture-label value is not an interaction, and the remainder of the paper
locates the step that fails.

The $O(G^6)$ residual of Eq.~(\ref{eq:area}) is disposed of by the earlier
argument rather than by this one. It produces an equilibrium force at fifth
post-Newtonian order, and Eq.~(\ref{eq:relabelaxi}) applies there as everywhere
else. With the density held at $\dA/4G$, the proportional identification, which
labels the points of each horizon so that the entropy density is uniform in the
labels at every separation (the cumulative-entropy labeling of
Sec.~\ref{sec:candidate}), returns for the first hole the contribution
$(Q_1/S_1)\,dS_1/d\rt=3G^6m_1^5m_2^2/1120\,\rt^7$ (Appendix~\ref{app:moebius}).
The label velocity $v=(3G^4m_1^3m_2/280\,\rt^5)\sin\theta$, fixed by
$\int(\partial_\theta T)\,v\,dS=(Q_1/S_1)\,dS_1/d\rt$, removes that contribution,
so the residual has no labeling-independent content either. Fifth post-Newtonian order is also where
conservative finite-size tidal effects first enter, and for black holes in four
dimensions the static Love numbers
vanish~\cite{BinningtonPoisson2009,DamourNagar2009,NoLove2025}. That is
consistent with the conclusion but not needed for it.

\begin{figure}[tbp]
\includegraphics[width=\columnwidth]{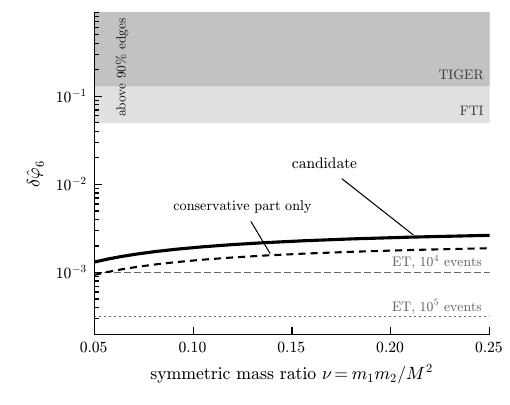}
\caption{The size of the candidate of Eq.~(\ref{eq:ratio}) if interpreted as a
physical modification: the fractional shift $\delta\hat\varphi_6$ of the third
post-Newtonian phasing coefficient that follows from
$\delta\varphi_6=-35\nu/3$ (solid), and from the conservative part alone,
$-25\nu/3$ (dashed). Shaded: the regions above the positive edges, $+0.05$
(FTI, lighter) and $+0.13$ (TIGER, darker), of the restricted $90\%$ GWTC-4.0
intervals, with the deviation taken common to all events~\cite{GWTC4TGR}.
Horizontal lines: forecast Einstein Telescope sensitivity after $10^4$ (dashed)
and $10^5$ (dotted) events~\cite{ET2025}. The curves are the content of a
prescription that this paper sets aside.}
\label{fig:pn}
\end{figure}

If the candidate were nevertheless interpreted as a physical modification, it
would be constrained by parametrized waveform tests. It shifts the 3PN phasing
coefficient by $\delta\varphi_6=-35\nu/3$, of which the conservative dynamics
supplies $-80\,\delta e_3=-25\nu/3$ and the change of the quadrupole flux at
fixed frequency the remaining $-10\nu/3$ (Appendix~\ref{app:pn}). As a fraction
of the general-relativistic value, the deviation
$\delta\hat\varphi_6\equiv\delta\varphi_6/\varphi_6^{\rm GR}$ used in
parametrized tests is $+2.6\times10^{-3}$ at $\nu=\tfrac14$ and
$+1.9\times10^{-3}$ at $\nu=0.1$. The GWTC-4.0 parametrized tests~\cite{GWTC4TGR}
report, for the flexible theory-independent (FTI) and Test Infrastructure for
General Relativity (TIGER) pipelines, restricted $90\%$ intervals, which take the deviation to be common to all
events, as in the earlier GWTC-3 analysis~\cite{GWTC3TGR}; their positive edges
are $+0.05$ (FTI) and $+0.13$ (TIGER). The candidate is common to all events
only approximately, its $\delta\hat\varphi_6$ ranging from $1.9$ to
$2.6\times10^{-3}$ as $\nu$ runs from $0.1$ to $\tfrac14$. Since it is positive,
the positive edges, which are one-sided $95\%$ bounds, are the relevant
comparisons, and it is smaller than them by factors of about 19 and 49 at
$\nu=\tfrac14$ and about 26 and 68 at $\nu=0.1$. The GWTC-5.0 tests likewise
report no departure from general relativity~\cite{GWTC5TGR}. A deviation of
this size would be within the forecast reach of the Einstein Telescope after
some $10^4$ catalog events~\cite{ET2025} (Fig.~\ref{fig:pn}). This is the
quantitative content of a hypothesis we have set aside, not a prediction of the
present work.

\section{The area dipole is a frame, not a shape}
\label{sec:isometry}

The puncture-label value cannot be an interaction, and Eq.~(\ref{eq:relabelaxi})
shows that its value depends on the labeling. This section shows where the value
comes from. Through fourth post-Newtonian order the candidate lies entirely in
the dipole of the area distribution, and that dipole is a property of the frame
in which the horizon is described rather than of the horizon.

\subsection{The multipole hierarchy}

For the axisymmetric data expand
\begin{align}
T(\theta)&=\sum_\ell T_\ell P_\ell(\cos\theta),\\
s(\theta)&\equiv\frac{dS}{d\Omega}=\sum_\ell s_\ell P_\ell(\cos\theta),
\end{align}
with $d\Omega$ the solid-angle element about the puncture; in this section and
in Appendices~\ref{app:relabel} and~\ref{app:tt}, $T_\ell$ and $s_\ell$ denote
these Legendre coefficients on a given horizon, not the temperatures of the two
holes. Orthogonality turns the contribution $F^{(i)}$ of the $i$th horizon to
Eq.~(\ref{eq:F}), $F=\sum_iF^{(i)}$, into
\begin{equation}
F^{(i)}=\sum_\ell\frac{4\pi}{2\ell+1}\,T_\ell\,\partial_{\rt}s_\ell .
\label{eq:forceLmain}
\end{equation}
The companion's multipole of order $\ell$ enters $T$ and $s$ with relative
amplitude $O(\eps^{\ell+1})$, so the term of order $\ell$ is of relative order
$\eps^{2\ell+1}$ in $F/F_{\rm N}$ for $\ell\ge1$. The monopole derivative
$\partial_{\rt}s_0$ first appears with the total-area correction of
Eq.~(\ref{eq:area}), at 5PN. The tidal quadrupole has $T_2=O(\eps^3)T_0$ and
$s_2=O(\eps^3)s_0$, so its first contribution is also at 5PN, and $\ell\ge3$
enters at 7PN; at leading order the $\ell=2$ coefficients are
$\kappa_{\ell=2}=-G^2m_1m_2/8\rt^3$ and
$(\dA_1/d\Omega)_{\ell=2}=G^5m_1^4m_2/\rt^3$. Neither depends on the quadrupolar
displacement of the horizon,
$\delta\ln\varrho_1=-\tfrac{10}{7}\,\eta\,P_2(\cos\theta)$ with
$\eta=G^3m_1^2m_2/16\rt^3$ (Appendix~\ref{app:isometry}), because $\kappa$ and
$\psi^4\rho^2$ are stationary under radial displacement at the Schwarzschild
throat; the displacement enters instead the shear (Sec.~\ref{sec:dynamics}) and
the area correction~(\ref{eq:area}). The dipole terms are
$T_1=-3Gm_2/32\pi\rt^2$ and, from
\begin{equation}
\frac{\dA_1}{d\Omega}=4G^2m_1^2
\left[1+\frac{G^2m_1m_2}{2\rt^2}\cos\theta+\cdots\right],
\label{eq:areadipmain}
\end{equation}
$s_1=G^3m_1^3m_2/2\rt^2$, and $(4\pi/3)\,T_1\,\partial_{\rt}s_1$ reproduces the
single-horizon term $G^4m_1^3m_2^2/8\rt^5$ of Eq.~(\ref{eq:Fpair}). The
temperature is lower, and the area density higher, on the side facing the
companion.

Direct multipole decomposition of the exact calculation confirms that the
complete 3PN and 4PN candidate is $\ell=1\times\ell=1$. The leading non-dipole
terms are known exactly: $\ell=0$ and $\ell=2$ first enter the single-horizon
term as $3G^6m_1^5m_2^2/1120\,\rt^7$ and $3G^6m_1^5m_2^2/80\,\rt^7$, so its
non-dipole fraction tends to $\tfrac{9}{28}(Gm_1/\rt)^2$. Summing the two
horizons, and using $m_1m_2(m_1^3+m_2^3)=\nu(1-3\nu)M^5$,
\begin{equation}
\frac{F_{\ell\neq1}}{F_{\rm N}}=\frac{9}{224}\,\nu(1-3\nu)\,\eps^5
+O(\eps^6),
\label{eq:F5PNmain}
\end{equation}
of which $\tfrac{3}{1120}$ comes from $\ell=0$ and $\tfrac{3}{80}$ from
$\ell=2$, and the non-dipole fraction of the total tends to
$\tfrac{9}{28}(1-3\nu)\eps^2$. A spectral horizon solver run at 35 to 45 digits
gives, for equal masses, non-dipole fractions of $6.9\times10^{-4}$,
$1.86\times10^{-4}$, and $4.83\times10^{-5}$ at $\rt=20$, $40$, and $80$ in units
$G=m_1=m_2=1$, unchanged to twelve digits when the Legendre truncation is varied
from 16 to 28 and approaching the prediction with $O(\eps)$ corrections; for
$m_1=2m_2$ the fraction reaches $0.98$ of the prediction at $\rt=53\,GM$. For
Brill--Lindquist data in puncture labels, therefore,
\begin{equation}
\boxed{\begin{gathered}
\text{3PN and 4PN are entirely }\ell=1\times\ell=1,\\
\text{genuine }\ell=0,2\text{ geometry first contributes at 5PN}.
\end{gathered}}
\label{eq:l1hiermain}
\end{equation}
On the physical slice the static transverse-traceless field modifies these
statements only through relabelings (Appendix~\ref{app:tt}). The companion's
potential and uniform field source no part of it, so that near each hole it
reduces at leading order to a constant tensor $\mathcal C_{ij}$, even under
reflection through the puncture, which relabels the horizon only in the
$\ell=2$ sector, while its part odd about the puncture is suppressed by a
further power of $Gm_i/\rt$. The candidate on the physical slice therefore
equals its Brill--Lindquist value through 3PN in every multipole, and the two
can differ at 4PN, through the odd part at $\ell=1$ and through $\mathcal C$ at
$\ell=2$. The first statement of Eq.~(\ref{eq:l1hiermain}) therefore refers to
Brill--Lindquist data in puncture labels, while the second holds on both slices
as a statement of order.

The question is thus not whether the horizon is round but whether its area
dipole belongs to its geometry.

\subsection{The area dipole is not intrinsic}

At linear order about a round sphere of areal radius $R_h$, a pure-trace
perturbation $\delta q_{AB}=f\,q_{AB}$ changes the Gaussian curvature by
\begin{equation}
\delta K=-\frac12\left(\Delta+\frac{2}{R_h^2}\right)f,
\label{eq:dKmain}
\end{equation}
where $\Delta$ is the Laplacian of the sphere, and any perturbation can be
brought to this form at $\ell=1$ by a tangential gauge choice. For an $\ell=1$
harmonic $\Delta f=-2f/R_h^2$, so
\begin{equation}
\delta K_{\ell=1}=0 .
\label{eq:dKl1main}
\end{equation}
No field equation is needed to see that the area-density dipole is not a
first-order intrinsic deformation of the two-geometry: an $\ell=1$ change of the
area density is generated by a conformal Killing field of the sphere.

The statement survives beyond linear order. Uniformization associates with any
smooth metric on $S^2$ a three-parameter family of conformally related unit
round metrics, related to one another by M\"obius transformations, and Hersch's
lemma~\cite{Hersch1970} shows that the family contains one for which
\begin{equation}
\boxed{\int_{S^2}Y^{\circ}_{1m}\,\dA=0,\qquad m=-1,0,1,}
\label{eq:balancedmain}
\end{equation}
where $Y^{\circ}_{\ell m}$ are its spherical harmonics. Ashtekar, Khera,
Kolanowski, and Lewandowski give an independent proof of existence, prove that
the balanced frame is unique up to rotations, using the positivity of the area
form, which holds here, and show that it inherits any axial Killing field of the
physical metric~\cite{AshtekarKheraKolanowskiLewandowski2022}. They present the
balancing condition as a conceptually simple gauge fixing rather than a
canonical one; we use only that it is fixed by the intrinsic geometry once the
area is balanced. For the axisymmetric Brill--Lindquist horizons the frame is
then unique up to rotations about the axis, and with $d\Omega^\circ$ its
solid-angle element the entropy density $s^\circ\equiv dS/d\Omega^\circ$ has
\begin{equation}
s^\circ_{1m}=0
\label{eq:s1zero}
\end{equation}
at every separation. This holds for the tidally distorted horizon as it is; it
does not require the physical horizon to be round, and the genuine quadrupolar
and higher distortions are retained.

\subsection{The force in the balanced frame}

The balanced frame can be constructed explicitly on each exact horizon. For an
axisymmetric cross-section $\rho=\varrho_1(\theta)$ the induced metric is
conformal to $(1+\varrho_1'^2/\varrho_1^2)\,d\theta^2+\sin^2\theta\,d\phi^2$,
with the prime denoting $\partial_\theta$, so the polar angle $\Theta$ of a
conformally related round metric satisfies
\begin{equation}
\ln\tan\frac{\Theta}{2}=\ln\tan\frac{\theta}{2}
+\int_0^\theta\Bigl[\sqrt{1+\varrho_1'^2/\varrho_1^2}-1\Bigr]
\frac{d\theta'}{\sin\theta'}+c ,
\label{eq:axiuniform}
\end{equation}
where the constant $c$ is the axial M\"obius parameter, fixed by
Eq.~(\ref{eq:balancedmain}), which here reads $\oint\cos\Theta\,\dA=0$.
Identifying points by $\Theta$ and differentiating the entropy density at fixed
$\Theta$ defines the balanced-frame force $F_{\rm bal}$. Its dipole term vanishes
identically, so the 3PN and 4PN terms of the puncture-label series,
Eq.~(\ref{eq:forceseries}), are absent, and
\begin{equation}
\frac{F_{\rm bal}}{F_{\rm N}}=\frac{9}{224}\,\nu(1-3\nu)\,\eps^5
\Bigl[1+O(\eps)\Bigr].
\label{eq:Fbalmain}
\end{equation}
The leading coefficient is that of Eq.~(\ref{eq:F5PNmain}), because the change
of frame mixes the dipole into the higher multipoles only at relative order
$\eps$. Like that of Eq.~(\ref{eq:F5PNmain}), it is a Brill--Lindquist value:
on the physical slice the static tidal response that conformally flat data
cannot represent can change it, though not its order (Appendix~\ref{app:tt}).
The spectral solver confirms this asymptote for mass ratios $1$ and $2$. For
equal masses $F_{\rm bal}/F_{\rm N}$ divided by the leading term of
Eq.~(\ref{eq:Fbalmain}) is $0.640$, $0.797$, $0.892$, and $0.944$ at $\rt=20$,
$40$, $80$, and $160$ in units $G=m_1=m_2=1$, while the same ratio for
$F_{\ell\neq1}$ is $0.658$, $0.808$, $0.898$, and $0.947$; for $m_1=2m_2$ the
balanced-frame ratio is $0.764$, $0.872$, and $0.933$ at $\rt=40$, $80$, and
$160$ in units $G=m_2=1$. All of these values are unchanged to the digits
quoted between Legendre truncations $16$ and $28$. The two quantities differ at
finite separation and share the leading coefficient. In the balanced frame
$s^\circ_1$ vanishes to working precision, and the same values of $F_{\rm bal}$
follow from Eq.~(\ref{eq:relabelaxi}) with the label velocity obtained from
Eq.~(\ref{eq:axiuniform}), the two evaluations agreeing to all quoted digits.

\subsection{Where the dipole comes from}

The Brill--Lindquist data show the origin of the dipole explicitly. Use
Cartesian coordinates $x^i$ centered on the first puncture, with $\rho=|x|$ and
the $z$ axis toward the companion. Expanding Eq.~(\ref{eq:psi}),
\begin{equation}
\psi=a+\frac{G\mu_1}{2\rho}+c_1z+O(\rho^2),\qquad
a=1+\frac{G\mu_2}{2\rt},\quad c_1=\frac{G\mu_2}{2\rt^2}.
\label{eq:nearfield}
\end{equation}
The constant $a$ is absorbed into the physical mass. The term $c_1z$ is the
uniform part of the companion's field, and it is generated by the special
conformal vector
\begin{equation}
\zeta^i=\beta\,(2zx^i-\rho^2\delta^i_z),\qquad \beta=\frac{c_1}{a}.
\label{eq:sct}
\end{equation}
Writing $\psi_0=a+G\mu_1/(2\rho)$, the terms proportional to $\mu_1$ cancel and
\begin{equation}
\Lie_\zeta(\psi_0^4\delta_{ij})=4\psi_0^3\,c_1z\,\delta_{ij},
\label{eq:lie}
\end{equation}
which exactly reproduces the linear companion field
(Appendix~\ref{app:isometry}).

The statement is exact, not only linear. The finite special conformal map
$\phi_\beta$ generated by Eq.~(\ref{eq:sct}) fixes the puncture, has flat
conformal factor $\Lambda=(1-2\beta z+\beta^2\rho^2)^{-1}$, and satisfies
$|\phi_\beta(x)|=\rho\,\Lambda^{1/2}$. It carries $\psi_0^4\delta_{ij}$
isometrically into $\tilde\psi^4\delta_{ij}$ with
\begin{align}
\tilde\psi&=\Lambda^{1/2}\,\psi_0\circ\phi_\beta
=\frac{G\mu_1}{2\rho}+\frac{a}{\sqrt{1-2\beta z+\beta^2\rho^2}}\nonumber\\
&=\frac{G\mu_1}{2\rho}+a\sum_{\ell\ge0}(\beta\rho)^\ell P_\ell(\cos\theta),
\label{eq:sctfinite}
\end{align}
the image of the constant $a$ being a point source at distance $1/\beta$. The
$\ell=0$ and $\ell=1$ terms coincide with those of Eq.~(\ref{eq:psi}) expanded
about the first puncture, so that
\begin{equation}
\psi=\tilde\psi+\sum_{\ell\ge2}\Bigl(\frac{G\mu_2}{2\rt^{\,\ell+1}}-a\beta^\ell\Bigr)
\rho^\ell P_\ell(\cos\theta)
\label{eq:sctremainder}
\end{equation}
exactly, the remainder being harmonic, regular at the puncture, and purely
$\ell\ge2$. Its coefficients are
$(G\mu_2/2\rt^{\ell+1})[1-(G\mu_2/2a\rt)^{\ell-1}]$. The constant and the
uniform field therefore change no invariant at any order in $c_1$, and the
leading remainder is the tidal quadrupole, of amplitude
$(G\mu_2/2\rt^3)(1-G\mu_2/2a\rt)$.

On the horizon, of coordinate radius $\rho_0=G\mu_1/2a$ at this order, the map
is a M\"obius boost of rapidity $\beta\rho_0=G^2\mu_1\mu_2/4a^2\rt^2$, which is
$G^2m_1m_2/4\rt^2$ at leading order. A boost of rapidity $\beta\rho_0$
multiplies the area element of the round sphere by
$(\cosh\beta\rho_0-\sinh\beta\rho_0\cos\theta)^{-2}\simeq1+2\beta\rho_0\cos\theta$,
which is why the rapidity is half the coefficient of the area dipole in
Eq.~(\ref{eq:areadipmain}). The boost found by numerical balancing approaches
this value: for equal masses it is $0.908$, $0.952$, $0.976$, and $0.988$ of it
at $\rt=20$, $40$, $80$, and $160$ in units $G=m_1=m_2=1$, unchanged between
Legendre truncations $16$ and $28$. The map induces the label velocity
$v=(G^2m_1m_2/2\rt^3)\sin\theta$, and by Eq.~(\ref{eq:relabelaxi}) the shift
$\int(\partial_\theta T)\,v\,dS=G^4m_1^3m_2^2/8\rt^5$ is the entire
single-horizon 3PN candidate (Appendix~\ref{app:moebius}). The area dipole is
thus the offset of the puncture labels from the horizon's own center-of-area
frame, and the candidate is the temperature-weighted rate at which that offset
changes with separation, equivalently the drift of the horizon's center of area
relative to the puncture. This statement is deliberately order specific: the
horizon acquires a genuine quadrupolar distortion at the next relative order,
and what Eq.~(\ref{eq:s1zero}) shows is that its area dipole remains frame
information rather than shape information.

The temperature dipole has the same origin, the uniform field seen through a
lapse normalized at infinity. It is not an intrinsic surface-gravity multipole of
a freely falling, body-adapted frame, and Sec.~\ref{sec:pnhorizon} makes that
distinction explicit using the Taylor--Poisson construction. Both dipoles
describe how the horizon looks from the puncture frame and from infinity rather
than the horizon itself, and in a freely falling frame the uniform field is
absent.

\subsection{What the balanced frame does and does not show}

The balanced frame does not make Eq.~(\ref{eq:W}) invariant at nonuniform $T$,
and Eq.~(\ref{eq:Fbalmain}) is not offered as a physical force.
Equation~(\ref{eq:relabel}) applies at every order. The cumulative-entropy
labeling of Sec.~\ref{sec:candidate} gives $F'=0$ through 4PN for every
temperature profile, since through that order the total area is separation
independent. At 5PN it returns the equilibrium term $(Q_i/S_i)\,dS_i/d\rt$,
which the label velocity of Sec.~\ref{sec:corroboration} removes, and the
construction of Sec.~\ref{sec:relabel} then gives the 5PN term any value,
including zero. What distinguishes the 3PN and 4PN terms is that the
identification fixed by the intrinsic geometry gives zero for them. A physical
value at 5PN would need a physical process, a thermodynamic path, and a balance
law, which the area density alone cannot supply.

\section{Dynamical transport}
\label{sec:dynamics}

An evolution supplies an identification, so one can ask directly whether gravity
moves area across the horizon in the way the candidate requires. The two
calculations of this section corroborate the conclusion; they do not replace the
relabeling identity, since each rests on a geometric choice of its own, and the
identity is what shows that some choice is needed.

If the puncture-label dipole of Eq.~(\ref{eq:areadipmain}) were carried along as
the separation changes at fixed physical masses, the area density would have to
change at the rate
\begin{equation}
\boxed{\begin{aligned}
\left.\partial_t\ln\frac{\dA}{d\Omega}\right|_{\ell=1,{\rm req}}
&=-\frac{G^2m_1m_2\,\dot{r}_{12}}{\rt^3}\cos\theta+\cdots\\
&=\frac{G^3Mm_1m_2}{\rt^5}\,t\cos\theta+\cdots,
\end{aligned}}
\label{eq:requiredmain}
\end{equation}
where the second form holds as the holes begin to fall from rest, with the
Newtonian estimate $\dot{r}_{12}=-GMt/\rt^2+\cdots$ near the time-symmetric
instant $t=0$. This rate and the transported rates below all vanish at $t=0$, as any rate must
at a turning point; an exact zero there is not evidence against a conservative
force, which has zero power at a turning point. The informative comparison is
between the first nonzero coefficients.

\subsection{The marginally trapped tube of the data}
\label{sec:mottmain}

Evolve the time-symmetric data with lapse $\mathcal N$ and zero shift. The
horizon is a strictly stable MOTS, the principal eigenvalue of its stability
operator being close to $1/4G^2m_1^2>0$, so it evolves into a smooth marginally
outer trapped tube (MOTT)~\cite{AnderssonMarsSimon2008}. With the null normals
of Sec.~\ref{sec:setup}, the tube is generated by
\begin{equation}
\begin{gathered}
V_t=\mathcal N\hat{u}+f\hat{N}=A\,k-B\,n,\\
A=\frac{\mathcal N+f}{2},\qquad B=\frac{f-\mathcal N}{2},
\end{gathered}
\label{eq:ABmain}
\end{equation}
where $f$ is the normal speed of the horizon within the slices. Keeping
$\theta_{(k)}=0$ along the tube fixes $f$. The variation of $\theta_{(k)}$, the
Gauss relation, and the vacuum constraint $R=0$ combine into
(Appendix~\ref{app:tube})
\begin{equation}
\boxed{(-2\Delta+R^{(2)})B=A\,|\sigma^{(k)}|^2,}
\label{eq:Beqmain}
\end{equation}
where $\Delta$ and $R^{(2)}$ are the Laplacian and scalar curvature of the MOTS
and $\sigma^{(k)}$ is the shear of $k$. Since $A=\mathcal N+B$, this is a linear
equation for $B$, whose operator $-2\Delta+R^{(2)}-|\sigma^{(k)}|^2$ is twice the
stability operator of the MOTS. The area element changes as
\begin{equation}
\boxed{\partial_t\ln\dA=-B\,\theta_{(n)},}
\label{eq:Btransportmain}
\end{equation}
where $\theta_{(n)}$ vanishes at $t=0$ and grows linearly in $t$; for a harmonic
lapse, which keeps the slicing maximal, its rate is
$\partial_t\theta_{(n)}=2\dot{K}_{\hat{N}\hat{N}}$, with $\dot{K}_{ij}$ the time
derivative of the extrinsic curvature of the slices. An independent
numerical reconstruction finds the solution of Eq.~(\ref{eq:Beqmain}) to agree
with $(f-\mathcal N)/2$ to relative errors of order $10^{-23}$ in the
high-precision runs.

The large-separation source can be obtained analytically. The leading tidal
quadrupole of Sec.~\ref{sec:isometry} gives
\begin{equation}
|\sigma^{(k)}|^2=
\frac{225G^4m_1^2m_2^2}{25088\,\rt^6}\sin^4\theta+O(\rt^{-7}),
\label{eq:sigclosedmain}
\end{equation}
and when $\mathcal N$ is constant the ratio $C\equiv B/A$ solves, at leading
order, $(-2\Delta+R^{(2)})C=|\sigma^{(k)}|^2$, which on the round throat gives
\begin{equation}
C=\frac{15G^6m_1^4m_2^2}{1568\,\rt^6}
\left(1-\frac{10}{49}P_2+\frac{1}{49}P_4\right)+O(\rt^{-7}),
\label{eq:Cclosedmain}
\end{equation}
an analytic check on the source and the inverse operator.

The tube depends on the foliation. Two lapses are natural, and both are maximal
at $t=0$, because the extrinsic curvature of the slice vanishes there and both
are harmonic: the lapse $\mathcal N=2/\psi-1$ of Eq.~(\ref{eq:chi}), which
defines the temperature weight, and the unit lapse $\mathcal N=1$, which
advances every end, the horizons included, at the asymptotic rate. Both tend to
$1$ at infinity, so in both $t$ is asymptotic time. Both tubes are nearly null,
since $B/A=C=O(\rt^{-6})$ for either lapse. On the MOTS of the first hole the
temperature lapse is
$\mathcal N=-Gm_2/2\rt-(G^2m_1m_2/4\rt^2)\cos\theta+\cdots$, small and negative,
so its tube advances slowly and, near the hole, the $t>0$ half lies to the past
of the $t=0$ surface and the $t<0$ half to the future; the coefficient of $t$
used below is common to both halves. The unit-lapse tube is future directed.

For both lapses the throat evolves, at leading order, as it would under a
constant lapse $\mathcal N_0$. For the unit lapse this is trivial, with
$\mathcal N_0=1$. For the temperature lapse it holds because near the hole
$2/\psi-1$ is the static lapse of the isolated throat, rescaled, plus the
constant $\mathcal N_0=1/a-1\simeq-Gm_2/2\rt$, with $a$ the constant of
Eq.~(\ref{eq:nearfield}); the static part vanishes on the horizon and
generates no evolution of the throat (Appendix~\ref{app:tube}). Both factors in
Eq.~(\ref{eq:Btransportmain}) then carry $\mathcal N_0$: $B\simeq\mathcal N_0C$,
and $\partial_t\theta_{(n)}=2\dot{K}_{\hat{N}\hat{N}}$ with
$\dot{K}_{ij}\simeq\mathcal N_0R_{ij}$, where $R_{ij}$ is the Ricci tensor of the
slice. The monopole of the transported rate therefore scales as
$\mathcal N_0^2$ and at leading order is
\begin{equation}
\left.\partial_t\ln\dA\right|_{\ell=0}
=\mathcal N_0^2\,\frac{15G^4m_1^2m_2^2}{3136\,\rt^6}\,t+\cdots,
\label{eq:mottmono}
\end{equation}
which is $R^{(2)}C_0\mathcal N_0^2\,t$ with $R^{(2)}=1/2G^2m_1^2$ the curvature
of the throat and $C_0$ the monopole of Eq.~(\ref{eq:Cclosedmain}). For the unit
lapse $\mathcal N_0=1$. For the temperature lapse
$\mathcal N_0^2=(1/a-1)^2\simeq G^2m_2^2/4\rt^2$, two further powers of the
separation. The dipole does not factor in this way: the temperature lapse has
its own dipole on the throat, $\mathcal N_1\cos\theta$ with
$\mathcal N_1=-G^2m_1m_2/4\rt^2$, which gives $\partial_t\theta_{(n)}$ the
dipole-to-monopole ratio $3\mathcal N_1/\mathcal N_0=3Gm_1/2\rt$ at leading
order and is absent for the unit lapse. The two transported dipoles therefore
still differ by two powers of the separation, $\rt^{-9}$ against $\rt^{-7}$,
but their ratio is not $\mathcal N_0^2$: at $\rt=160$ it is
$1.43\times10^{-4}$, about $3.7\,\mathcal N_0^2$. An exact order-by-order solution of
Eq.~(\ref{eq:Beqmain}) for the temperature lapse gives
\begin{align}
\left.\partial_t\ln\dA\right|_{\mathcal N=2/\psi-1}
&=\frac{15\,G^6m_1^2m_2^4}{12544\,\rt^8}\,t\nonumber\\
&\quad+\frac{9495\,G^7m_1^3m_2^4}{3995264\,\rt^9}\,t\cos\theta+\cdots,
\label{eq:mottclosed}
\end{align}
whose monopole is Eq.~(\ref{eq:mottmono}). Direct numerical solution confirms
Eq.~(\ref{eq:mottclosed}) to separations of $\rt=2560$ ($\mu_1=1$, $\mu_2=2$,
$G=1$): Richardson extrapolation from $\rt=160$, $640$, and $2560$ of the ratios
of the numerical monopole and dipole to the closed forms gives $0.999997$ for
the monopole and $0.999995$ for the dipole, identically at the resolutions
$(12,40)$ and $(16,56)$, the first number being the Legendre truncation and the
second the number of quadrature nodes. For the unit lapse the transported
dipole scales as $t\,\rt^{-7}$; between $\rt=80$ and $160$ the local exponents
of the numerical monopole and dipole are $-5.91$ and $-6.89$, approaching $-6$
and $-7$. At $\rt=160$ the ratio of the two numerical monopoles is
$3.858\times10^{-5}$, against $\mathcal N_0^2=(1/a-1)^2=3.858\times10^{-5}$.

The transported dipoles fall far below Eq.~(\ref{eq:requiredmain}). For
$\mu_1=1$, $\mu_2=2$, the unit-lapse dipole is $2.1\times10^{-6}$ of the
required one at $\rt=20$, where $\rt\simeq6.5\,GM$, and $6.1\times10^{-8}$ at
$\rt=160$; for the temperature lapse the ratios are $1.8\times10^{-8}$ and
$8.7\times10^{-12}$. These values are identical at the two resolutions
$(18,64)$ and $(24,96)$. Runs at the mass ratios $\mu_2/\mu_1=1$ and
$\tfrac12$ test the mass dependence of Eq.~(\ref{eq:mottclosed}) in the same
way: the Richardson-extrapolated ratios lie within $2\times10^{-6}$ of unity for
both multipoles, identically at the two resolutions, and the unit-lapse local
exponents between $\rt=80$ and $160$ are $-5.96$ and $-6.94$. Thus
\begin{equation}
\boxed{\begin{gathered}
\text{required (puncture-label) dipole rate}\sim\rt^{-5},\\
\text{transported dipole, }\mathcal N=1:\ \sim\rt^{-7},\\
\text{transported dipole, }\mathcal N=2/\psi-1:\ \sim\rt^{-9}.
\end{gathered}}
\label{eq:mottgapmain}
\end{equation}
The comparison does not convert a time rate into a conservative force, and it
does not assert uniqueness under tangential additions to $V_t$. Marginally
trapped tubes depend on the foliation, so we use them only to test whether these
evolutions reproduce the candidate. They do not.

\subsection{The tidally perturbed post-Newtonian horizon}
\label{sec:pnhorizon}

The standard matched-asymptotic construction of a black hole in a tidal
environment supplies an independent check in the full dynamics. Taylor and
Poisson choose a body-adapted frame for a nonrotating black hole in a
post-Newtonian external spacetime and remove all mass dipoles, the intrinsic one
and the tidal one; the vacuum even-parity $\ell=1$ Schwarzschild perturbation is
gauge~\cite{TaylorPoisson2008}. The uniform part of the companion's field is
therefore absent in that frame. Poisson and Vlasov extend the local tidal metric
through fourth tidal order, and Poisson and Corrigan give the higher tidal
moments explicitly for a
binary~\cite{Poisson2005,TaylorPoisson2008,PoissonVlasov2010,PoissonCorrigan2018}.
Set $G=c=1$ in this subsection. The physical tidal hierarchy is
\begin{equation}
\cE_{ab}\sim\frac{m_2}{b^3},\qquad
\cE_{abc}\sim\frac{m_2}{b^4},\qquad
\cB_{ab}\sim\frac{m_2V}{b^3},
\label{eq:tidalhiermain}
\end{equation}
where $\cE_{ab}$, $\cE_{abc}$ and $\cB_{ab}$ are the electric quadrupole,
electric octupole, and magnetic quadrupole tidal moments, $b$ is the orbital
separation, and $V$ the relative orbital speed. The environment thus begins at
$\ell=2$, not $\ell=1$. We write $\phi$ for the orbital phase, $\mathbf n$ for
the unit vector from the hole to its companion, and $\Omega^a$ for the radial
unit vector labeling a point of the horizon two-sphere.

We use two Poisson--Vlasov results. In angular coordinates constant along the
horizon generators [Eq.~(4.4) of Ref.~\cite{PoissonVlasov2010}],
\begin{equation}
\sqrt{\det q_{AB}}=4m_1^2\sin\theta\,[1+O(5)],
\label{eq:PVdetmain}
\end{equation}
although the horizon already carries genuine quadrupolar, octupolar, and
bilinear tidal deformations, so the generator-comoving area measure has no
$\ell=1$ moment through fourth tidal order. We have checked the $\ell=1$ part of
this result independently: at fourth order the $\ell=1$ parts of a trace term
and of an electric--magnetic cross term in the determinant cancel. Here $O(n)\equiv O[(m_1/\mathcal R)^n]$,
with $\mathcal R$ the radius of curvature of the tidal field. Poisson and Vlasov
find the expansion of the generators of the null surface $r=2m_1$ to vanish
through fourth tidal order, so that the surface is foliated by marginally outer
trapped cross-sections, and to that order they describe it as an isolated
horizon; the shear of Eq.~(\ref{eq:PVshearmain}) below is nonzero but enters the
expansion only quadratically, beyond that order. The statements of this
subsection therefore apply to the quasi-local thermodynamic surface adopted in
Sec.~\ref{sec:setup}. The surface coincides with the event horizon only under
the additional assumption that the tidal perturbation is eventually switched
off. The surface gravity, normalized on the generators, is
\begin{equation}
\begin{aligned}
\kappa=\frac{1}{4m_1}\Bigl[1
&+\frac{16}{3}m_1^3\dot{\cE}_{ab}\Omega^a\Omega^b
+\frac{32}{9}m_1^4\ddot{\cE}_{ab}\Omega^a\Omega^b\\
&+\frac{8}{9}m_1^4\dot{\cE}_{abc}\Omega^a\Omega^b\Omega^c\\
&-\frac{16}{225}m_1^4
  \bigl(\cE_{ab}\cE^{ab}+\cB_{ab}\cB^{ab}\bigr)
+O(5)\Bigr],
\end{aligned}
\label{eq:PVkappamain}
\end{equation}
with overdots denoting derivatives with respect to the advanced time $v$, which
is unrelated to the label velocity $v$ of Sec.~\ref{sec:candidate}. The
fourth-order terms are $\ell=2$, $3$, and $0$, so the body-adapted surface
gravity has no $\ell=1$ moment through fourth tidal order; it varies over the
horizon only through time derivatives of the tidal moments and is uniform for a
stationary tide~\cite{PoissonVlasov2010}. This $\kappa$ is normalized in the
generator frame, not at infinity as the Brill--Lindquist weight of
Eq.~(\ref{eq:Tlocal}) is, and the comparison is consistent with the
Brill--Lindquist temperature dipole being a redshift and frame effect rather
than an intrinsic multipole of the horizon's surface gravity.

The horizon shear begins as
\begin{equation}
\sigma_{AB}=-\frac43m_1^4
(\dot{\cE}^{q}_{AB}+\dot{\cB}^{q}_{AB})
-\frac4{15}m_1^5
(\dot{\cE}^{o}_{AB}+\dot{\cB}^{o}_{AB})+\cdots,
\label{eq:PVshearmain}
\end{equation}
where, with $\Omega^a_A=\partial\Omega^a/\partial\theta^A$,
$\cE^q=\cE_{ab}\Omega^a\Omega^b$ and $\cE^o=\cE_{abc}\Omega^a\Omega^b\Omega^c$,
\begin{align}
\cE^q_{AB}&=2\cE_{ab}\Omega^a_A\Omega^b_B+\cE^q\Omega_{AB},
\nonumber\\
\cE^o_{AB}&=2\cE_{abc}\Omega^a_A\Omega^b_B\Omega^c+\cE^o\Omega_{AB},
\label{eq:PVtensors}
\end{align}
$\Omega_{AB}$ is the metric of the unit sphere, and $\cB^q_{AB}$, $\cB^o_{AB}$
are the odd-parity counterparts. In this normalization $\cE^o_{AB}$ is one third
of the tracefree part of $D_AD_B\cE^o$, and the quadrupole term reproduces the
standard heating rate
$\dot{m}_1=(16/45)m_1^6(\dot{\cE}_{ab}\dot{\cE}^{ab}+\dot{\cB}_{ab}\dot{\cB}^{ab})$~\cite{Poisson2004};
these tensor-potential normalizations agree with Table~I of
Ref.~\cite{PoissonVlasov2010}.

The expansion $\vartheta$ of the horizon generators obeys
\begin{equation}
\partial_v\vartheta=\kappa\vartheta-\frac12\vartheta^2-\sigma_{AB}\sigma^{AB},
\label{eq:PVraymain}
\end{equation}
so area production starts quadratically in the tidal shear. For slow tides its
quasi-stationary solution is
$\vartheta=\sigma_{AB}\sigma^{AB}/\kappa_0+O(m_1\partial_v\sigma^2)$ with
$\kappa_0=1/4m_1$; under the switch-off assumption that identifies this null
surface with the event horizon, this is the teleological solution selected by
the future boundary condition.

For a general planar Newtonian orbit, now with overdots denoting orbital-time
derivatives, the radial dipole of the squared shear obtained from the
Poisson--Vlasov tensor harmonics is
\begin{align}
[\sigma^2]^{\rm rad}_{\ell=1}
={}&\frac{128}{35}\frac{m_1^5m_2^2}{b^7}
\left(\dot\phi^2+\frac{3\dot{b}^2}{b^2}\right)\notag\\
&\mp\frac{128}{5}\frac{m_1^4m_2^2\dot{b}\dot\phi^2}{b^6}+\cdots,
\label{eq:generalradialmain}
\end{align}
where the first group is electric quadrupole--octupole ($E_2E_3$) interference
and the second electric--magnetic quadrupole ($E_2B_2$) interference, whose
sign depends on the magnetic-tidal convention. Appendix~\ref{app:pvdipole}
gives the tidal moments and the projection. The coefficient-level calculation
has been checked independently from the Poisson--Vlasov tensor-potential
definitions and against their integrated-heating normalization. At fixed
physical masses the changing puncture-label dipole would require
\begin{equation}
\left.\partial_t\ln\frac{\dA}{d\Omega}\right|_{\ell=1,{\rm req}}
=-\frac{m_1m_2\dot{b}}{b^3}\,(\mathbf n\cdot\boldsymbol\Omega),
\label{eq:PNrequiredmain}
\end{equation}
which is the first form of Eq.~(\ref{eq:requiredmain}) with $\dot\rt=\dot{b}$. The part of the
physical expansion linear in $\dot{b}$ then gives the magnitude ratio
\begin{equation}
\boxed{\begin{aligned}
\left|\frac{\vartheta^{E_2B_2}_{\ell=1,{\rm rad}}}
{\partial_t\ln(\dA/d\Omega)|_{\ell=1,{\rm req}}}\right|
&=\frac{512}{5}\frac{m_1^4m_2M}{b^6}\,\Xi_\phi,\\
\Xi_\phi&\equiv\frac{b^3\dot\phi^2}{M}.
\end{aligned}}
\label{eq:sixpowersmain}
\end{equation}
For a Keplerian orbit of eccentricity $e$, $\Xi_\phi=1+e\cos(\phi-\phi_{\rm p})$
with $\phi_{\rm p}$ the phase at pericenter, so for generic nonradial motion it is
$O(1)$: the physical radial transport linear in $\dot{b}$ is suppressed by six
powers of $M/b$ relative to the rate the puncture-label dipole requires. No
``9PN force'' interpretation is implied, since the numerator is a local
area-production rate.

For head-on infall $\dot\phi=0$ and the $E_2B_2$ term vanishes. The surviving
$E_2E_3$ contribution gives
\begin{equation}
\vartheta_{\ell=1}^{\rm head\mbox{-}on}=
\frac{1536}{35}\frac{m_1^6m_2^2\dot{b}^2}{b^9}
(\mathbf n\cdot\boldsymbol\Omega)+\cdots,
\label{eq:headonthetamain}
\end{equation}
so its ratio to Eq.~(\ref{eq:PNrequiredmain}) has magnitude
$(1536/35)m_1^5m_2|\dot{b}|/b^6$, that is $O[(M/b)^{13/2}]$ at free-fall speeds
and zero at release. On a circular orbit the $\dot{b}$-independent $E_2E_3$
dipole instead scales as $b^{-10}$, but the puncture-label dipole is stationary
there and requires no transport; that term is a genuine physical horizon dipole
with no counterpart in the puncture-label construction. These results are consistent with the
standard fact that horizon absorption by nonspinning holes begins at 4PN
relative order~\cite{PoissonSasaki1995,TaylorPoisson2008}, but the observables
and the power counting differ, so that agreement is qualitative corroboration
rather than a conversion rule between orders.

\subsection{Trapping and event horizons}
\label{sec:sehbridge}

The two calculations should remain independent. Slowly evolving trapping
horizons and event horizons have closely related near-equilibrium flux
laws~\cite{BoothFairhurst2007}, but their local transport fields need not
coincide. Booth's near-horizon construction~\cite{Booth2013} makes the
distinction explicit. For the nearby null surface, the induced metric and area
element agree with those of a slowly evolving trapping horizon at leading order,
but the local expansions satisfy
\begin{equation}
\kappa_V\theta_{(V)}-\kappa^{(E)}_k\theta^{(E)}_{(k)}
\simeq D_B(D^BC-2C\widetilde\omega^B),
\label{eq:boothdivmain}
\end{equation}
in the notation of Ref.~\cite{Booth2013}, with his outgoing null normal written
$k$, $V$ the evolution vector and $C$ the expansion parameter of the trapping
horizon, $\widetilde\omega^B$ its rotation one-form, and the superscript $(E)$
marking the null surface. Booth's $C$ is our $B/A$, and his $V$ our $V_t$
normalized to $A=1$. The right-hand side is a tangential divergence: it can
alter local multipoles when $C$ is nonuniform, while integrating to zero on a
closed cross-section. That is Eq.~(\ref{eq:relabel}) in a dynamical setting,
with two horizons matched point by point. For the Brill--Lindquist tube the
point is not academic, since $B/A$ acquires a dipole at the order beyond
Eq.~(\ref{eq:Cclosedmain}).

Booth's construction moreover starts from a future outer trapping horizon, with
$\theta_{(k)}=0$ and strictly $\theta_{(n)}<0$. The Brill--Lindquist $t=0$ leaf
instead has $\theta_{(n)}=0$ by time symmetry, and its first evolution
coefficient is nonzero, so it lies on the boundary of the slowly evolving class
rather than within it. The coefficient of $t$ in the transport of the
Brill--Lindquist tube therefore cannot be identified, through that theorem, with
the coefficient of the post-Newtonian null horizon, which is the event horizon
under the switch-off assumption. The $\rt^{-7}$ and $\rt^{-9}$ tube results and
the tidal null-horizon result are complementary checks rather than the same
number in two descriptions.

\section{Uniform temperature and the first law}
\label{sec:uniform}

By Sec.~\ref{sec:relabel}, and in the axisymmetric form of
Eq.~(\ref{eq:relabelaxi}), the value of Eq.~(\ref{eq:F}) is an invariant of the
configuration only if
\begin{equation}
\partial_\theta T=0 .
\label{eq:uniform}
\end{equation}
When the temperature weight is uniform the shift vanishes for every relabeling,
$T$ comes out of the integral, and
\begin{equation}
\sum_j\int T_j\,\partial_{\rt}(dS_j)=\sum_jT_j\,\partial_{\rt}S_j ,
\label{eq:uniformforce}
\end{equation}
which involves only totals; here $T_j$ is the temperature of the $j$th hole.

A uniform surface gravity is what the zeroth law delivers. It holds on
stationary Killing horizons and, with the appropriate horizon structure, on
isolated horizons, and a static distorted hole held in place by external matter
has constant $\kappa$~\cite{GerochHartle1982}. Stationarity is therefore a
sufficient mechanism for Eq.~(\ref{eq:uniform}); we do not claim that it is
necessary, and nothing below requires that. What matters is the direction we do
have. The Brill--Lindquist holes are unsupported and momentarily at rest, so
$\xi$ is the evolution vector of the maximal slicing and not a Killing field;
out of equilibrium the surface gravity associated with such an evolution need
not be constant~\cite{BoothFairhurst2004}, and by Eq.~(\ref{eq:kappaexp}) it is
not. The stand-alone transport term therefore has no labeling-independent
content there. The criterion does not assert that a proper stretched-horizon
temperature must also be nonuniform, nor that a nonuniform temperature can never
take part in a physical local work law: if the dynamics supplies a material
velocity, a generator flow, a membrane current, or another physical connection,
that structure can define the transport. Equation~(\ref{eq:relabel}) says only
that the stand-alone functional built from the unlabeled family, $T$, and $dS$
does not.

The question is then what the construction returns when
Eq.~(\ref{eq:uniform}) does hold. Two black holes cannot be in static
equilibrium in vacuum, so something must hold them apart.

\subsection{A consequence of the first law}

Suppose a static configuration of black holes is maintained by some external
agency, and that its first law has the form
\begin{equation}
d\mathcal H=\sum_iT_i\,dS_i+\sum_a\Phi_a\,d\mathcal Q_a-\delta W_{\rm ext},
\label{eq:genfirstlaw}
\end{equation}
where $\mathcal H$ is the total mass, $\mathcal Q_a$ are the conserved charges
with potentials $\Phi_a$, and $\delta W_{\rm ext}$ is the term conjugate to the
agency's intensive variable. Along a one-parameter variation with
$d\mathcal H=0=d\mathcal Q_a$, Eq.~(\ref{eq:genfirstlaw}) reduces to
\begin{equation}
\sum_iT_i\,dS_i=\delta W_{\rm ext} .
\label{eq:theorem}
\end{equation}
Equation~(\ref{eq:genfirstlaw}) is an identity on the parameter space rather than
a relation along a chosen path, so the conclusion may be stated without choosing
one: where the construction has labeling-independent content, it returns
\begin{equation}
\sum_iT_i\,dS_i=d\mathcal H-\sum_a\Phi_a\,d\mathcal Q_a+\delta W_{\rm ext}
\label{eq:everypath}
\end{equation}
on every path, and dividing by the parameter of the variation gives a derivative
of the form appearing in Eq.~(\ref{eq:F}). The construction is the first law
rearranged and carries no more than the first law does; along a path of fixed
mass and charges it returns the work term of whatever maintains the
configuration. The argument uses only the form of Eq.~(\ref{eq:genfirstlaw}),
which is how the first law is written for static pairs of black holes held apart
by a conical defect, with and without charge~\cite{KrtousZelnikov2020,Appels2017},
and the conclusion has exactly that scope; we do not assert it for entropic
constructions in general.

\subsection{The Israel--Khan solution}

The statement can be seen in a case where every quantity is known in closed
form. The Israel--Khan solution describes two collinear Schwarzschild black holes
held apart by a conical singularity on the axis between
them~\cite{IsraelKhan1964,BachWeyl1922}, which acts as a strut. The spacetime is
static, the horizons are Killing horizons, and the zeroth law applies. The holes
are represented by rods on the axis of lengths $\lambda_i=2GM_i$, with $M_i$
their masses, separated by a coordinate gap $\Delta z$; we use $\lambda_i$
rather than $\mu_i$ because these are lengths. The surface gravity and area of
the first hole are~\cite{CostaPerry2000}
\begin{equation}
\kappa_1=\frac{1}{2\lambda_1}\frac{\Delta z+\lambda_1}{\Delta z+\lambda_1+\lambda_2},
\quad
\A_1=4\pi\lambda_1^2\frac{\Delta z+\lambda_1+\lambda_2}{\Delta z+\lambda_1},
\label{eq:IK}
\end{equation}
with $\kappa_1$ independent of position on the horizon, as the zeroth law
requires. Equation~(\ref{eq:uniform}) is satisfied and the construction is
unambiguous.

Our Smarr relation reappears exactly. The separation dependence cancels between
the factors of Eq.~(\ref{eq:IK}),
\begin{equation}
\kappa_1\A_1=2\pi\lambda_1=4\pi GM_1\quad\text{identically},
\label{eq:IKsmarr}
\end{equation}
so that $Q_1=\kappa_1\A_1/8\pi G=M_1/2$ with no expansion. This is
Eq.~(\ref{eq:Qexact}) in a different spacetime, obtained by a different route,
with a strut present. Equivalently $\mathcal H=2\sum_iT_iS_i$, the Smarr formula
for this system~\cite{KrtousZelnikov2020}, which carries no strut term because
the Smarr relation follows from Euler scaling and the tension is unchanged under
a rescaling of length.

For this system the first law is
$d\mathcal H=T_1\,dS_1+T_2\,dS_2-L\,d\tau$, with $\tau$ the strut tension and $L$
the conjugate thermodynamic length, equal to the strut worldsheet area per unit
Killing time~\cite{KrtousZelnikov2020,Appels2017}. Since
$\mathcal H=(\lambda_1+\lambda_2)/2G$, the path of fixed rod lengths realizes
$d\mathcal H=0$, and Eq.~(\ref{eq:theorem}) reads
\begin{equation}
\sum_i T_i\frac{\partial S_i}{\partial\Delta z}
=L\,\frac{\partial\tau}{\partial\Delta z}.
\label{eq:strutid}
\end{equation}
Taking $\kappa_i$ and $\A_i$ from Eq.~(\ref{eq:IK}) and the strut variables from
Ref.~\cite{KrtousZelnikov2020}, both sides equal
\begin{equation}
-\frac{\lambda_1\lambda_2\,(2\Delta z+\lambda_1+\lambda_2)}
{4G(\Delta z+\lambda_1)(\Delta z+\lambda_2)(\Delta z+\lambda_1+\lambda_2)}
\label{eq:strutvalue}
\end{equation}
identically, with zero residual (Appendix~\ref{app:strut}). The relation between
the variables of the two sources is fixed by their own rod geometry and is not
supplied by us.

What the entropic construction returns here is thus the thermodynamic work term
associated with the strut. We avoid calling it the work done against the strut,
because the pairing is between $L$ and the tension rather than between a force
and a displacement, and the sign is accordingly opposite to the ordinary
mechanical analogy~\cite{KrtousZelnikov2020}. It is in any case not the
gravitational attraction, and not merely by a normalization: at large separation
$\tau\to GM_1M_2/\Delta z^2$, which is the Newtonian force, while
$\tau\sim\Delta z^{-2}$ and $L\sim\Delta z$ give
$L\,\partial_{\Delta z}\tau\to-2\tau$. We do not connect this factor to the Komar
factor of Eq.~(\ref{eq:Qexact}).

\subsection{The thermodynamic path}
\label{sec:path}

Equation~(\ref{eq:theorem}) holds along a path of fixed total mass, and the
Brill--Lindquist calculation of Sec.~\ref{sec:candidate} does not follow such a
path: there the individual masses $m_i$ are held fixed and the total energy
$E_{\rm ADM}=\sum_i\mu_i=M-E_b$ varies with separation. The two prescriptions
must be compared explicitly.

Both correspondences are ours. Equation~(\ref{eq:Qexact}) gives $2Q_i=\mu_i$, so
$\mu_i$ is the mass variable associated with this Komar-normalized horizon
charge, and $E_{\rm ADM}=\sum_i\mu_i$; it is therefore the natural analog of the
fixed-$\lambda_i$ mass coordinate of Israel--Khan, although we do not call it a
Komar mass, since the data admit no Killing field. By the rigidity of
Sec.~\ref{sec:rigid}, $m_i$ and the irreducible mass coincide through five
relative orders, so holding $m_i$ fixed is the fixed-area path through the orders
relevant to Eq.~(\ref{eq:Fpair}).

The total-area part of the construction can be tracked in these variables.
Because $T$ is not uniform, $Q_i=\int T\,dS$ is not a product; define instead the
entropy-weighted mean $\bar{T}_i\equiv Q_i/S_i$, for which
$\bar{T}_i\,dS_i=Q_i\,d\ln S_i$ exactly. With $Q_i=\mu_i/2$ and $S_i=4\pi Gm_i^2$
this gives
\begin{equation}
\sum_i\bar{T}_i\,dS_i=\sum_i\frac{\mu_i}{m_i}\,dm_i
\label{eq:blcontent}
\end{equation}
through the area-rigid orders. It vanishes identically at fixed $m_i$, and at
fixed $\mu_i$ it gives $-G\mu_1^2\mu_2/2m_1\rt^2$ for the first hole and,
summed over the two horizons, $-G\mu_1\mu_2/\rt^2$ at leading order, which is
the Newtonian-scale term of Ref.~\cite{vanPutten2012}, recovered also in the
rebuild of Appendix~\ref{app:rebuild}. The same total-area structure therefore
appears in both calculations, the fixed-$\mu_i$ term being the analog, at the
level of the entropy-weighted mean temperature, of the horizon term obtained
along the fixed-mass Israel--Khan path. Their status differs, however, since
only the latter arises in a stationary configuration with uniform horizon
temperature, where the local construction is not relabeling dependent at all.

There is a thermodynamic way to state the fixed-$\mu_i$ result. Since
$E_{\rm ADM}=\mu_1+\mu_2$, holding the $\mu_i$ fixed is a path of constant total
energy, and at leading order
\begin{equation}
\sum_i\bar{T}_i\,\frac{dS_i}{d\rt}\Big|_{\mu_i}=-\frac{G\mu_1\mu_2}{\rt^2}
=-\Bigl(\frac{\partial E_{\rm ADM}}{\partial\rt}\Bigr)_{S_i}.
\label{eq:fixedmuE}
\end{equation}
This is the relation $T(\partial S/\partial r)_E=-(\partial E/\partial r)_S$
that holds identically for any first law of the form $dE=T\,dS-F\,dr$: along a
path of constant energy the entropy derivative restates the energy gradient at
fixed entropy rather than defining a new interaction. The identification stops
at leading order. Beyond it the Noether-charge mean weight is not the
thermodynamic derivative of the ADM energy,
\begin{align}
\left(\frac{\partial E_{\rm ADM}}{\partial S_1}\right)_{S_2,\rt}
&=\frac{1-Gm_2/\rt}{8\pi Gm_1}+\cdots,\nonumber\\
\bar{T}_1=\frac{Q_1}{S_1}
&=\frac{1-Gm_2/2\rt}{8\pi Gm_1}+\cdots,
\label{eq:firstlawmismatch}
\end{align}
so the exact Smarr identity of Sec.~\ref{sec:smarr} does not furnish a first law
for the binary with this weight. The fixed-$\mu_i$ Newtonian term is therefore a
restatement of the ordinary interaction energy at leading order, not an
additional horizon force.

For Israel--Khan, the fixed-$\lambda_i$ path keeps the total mass constant, and
Eq.~(\ref{eq:everypath}) gives Eq.~(\ref{eq:strutid}). Along the fixed-area path
the construction vanishes identically, $dS_i$ being zero, and the content of that
path is instead carried by the total mass: solving Eq.~(\ref{eq:IK}) for
$d\lambda_i/d\Delta z$ at fixed $\A_i$,
\begin{equation}
\sum_iT_i\frac{\partial S_i}{\partial\Delta z}=0,
\qquad
\frac{d\mathcal H}{d\Delta z}=\frac{2GM_1M_2}{\Delta z^2}+\cdots ,
\label{eq:pathB}
\end{equation}
which is asymptotically twice the Newtonian attraction, and the first law closes
with zero residual (Appendix~\ref{app:strut}). At leading order, under the
weak-field matching $\mu_i\leftrightarrow M_i$ and $\rt\leftrightarrow\Delta z$,
which is not an exact correspondence between the two spacetimes, each of these
Israel--Khan quantities is twice its Brill--Lindquist counterpart
(Table~\ref{tab:ikbl} of Appendix~\ref{app:strut}).

The value of Eq.~(\ref{eq:F}) thus depends on the thermodynamic path as well as
on the labeling, and on neither path is it a force between the holes. That is not
an accident of the two paths chosen: by Eq.~(\ref{eq:everypath}) the construction
is the first law rearranged and returns
$d\mathcal H-\sum_a\Phi_a\,d\mathcal Q_a+\delta W_{\rm ext}$ whatever path is
taken, which for the neutral system here is $d\mathcal H+\delta W_{\rm ext}$.

\section{Discussion}
\label{sec:discussion}

It is worth separating what has and has not been shown. The
Bekenstein--Hawking law and the first law are untouched, and nothing here shows
that horizon entropy has no density. What we show is that a density, even
supplemented by a nonuniform temperature normalized at infinity, does not by
itself determine a stand-alone entropy-transport contribution to mechanical
work, because differentiating a local measure requires a connection between
neighboring horizon cross-sections. Equation~(\ref{eq:relabel}) is the exact
statement of that obstruction: the stand-alone transport functional is invariant
under the admitted relabelings only when its weight is uniform, unless additional
physical transport structure is supplied. Localizing horizon entropy,
transporting it, and giving the transport a mechanical meaning are three
distinct steps: the first is an assumption about the density, the second needs
an identification of horizon points, and the third needs a thermodynamic path
and a balance law.

The two-black-hole example resolves a puzzle. A local reading of the area law,
weighted by a physically normalized temperature, produces a 3PN repulsion that
the post-Newtonian dynamics of general relativity does not contain. The
resolution is that the construction never had a value of its own. In
Brill--Lindquist data its first two nonzero post-Newtonian orders are entirely
$\ell=1\times\ell=1$, and the area dipole is removed in the balanced conformal
frame of Eq.~(\ref{eq:balancedmain}); the genuine quadrupole is present at
relative order $\eps^3$ but belongs to a different multipole sector and first
contributes at 5PN. The dynamical constructions, which do supply
identifications, agree.

The same freedom appears whenever local horizon rates are compared across
surfaces, as in Booth's comparison of trapping and event horizons
(Sec.~\ref{sec:sehbridge}). The full charge variation
$dQ=d\!\int T\,dS$ likewise remains invariant, because the change in the weight
compensates the change in the transport term, and when the weight is uniform the
first law fixes what the construction returns. The thermodynamic path and the
local identification are logically distinct choices, and both must be made.

Any account in which information falling into a black hole is recorded on
localized patches of the horizon must therefore specify a rule for transporting
those patches, and that rule is not contained in the area law. In a genuinely
dynamical spacetime the evolution itself may supply it, as the generators of an
evolving horizon do. This is a requirement on such accounts, not an argument
against them.

A stretched horizon gives a useful counterpoint. At fixed proper cutoff
$\rho_*$, $T_{\rm prop}\simeq(2\pi\rho_*)^{-1}$ can be approximately uniform, so
purely proper local work is simply $T_{\rm prop}\,dS$ and is insensitive to
relabeling. This does not yet define work measured at infinity. A local energy
increment is redshifted by the lapse, and in the stationary local-Rindler limit
its asymptotic weight is $\mathcal NT_{\rm prop}=\kappa/2\pi$. Varying the
corresponding integrated charge therefore produces both the entropy-transport
term and the variation of the redshift, the two terms of
Eq.~(\ref{eq:Qsplitmain2}), and only their sum is invariant. A universal
microscopic cutoff temperature can thus evade the local form of the obstruction,
but obtaining an asymptotic force still requires structure that specifies how
local horizon energy and momentum contribute to the asymptotic Hamiltonian. A
membrane stress tensor, a quasi-local energy prescription, or a microscopic
transport law could provide it; none is contained in the area law.

None of this says that black holes exert no entropic influence on one another.
Tidal heating raises the horizon area of each hole at a rate proportional, within
general relativity, to the squared shear of its
generators~\cite{Hartle1973,Thorne1986,Poisson2004}; on the quasi-local surface
adopted here that statement is the dynamical-horizon flux law rather than
Hawking's theorem, which concerns the event horizon~\cite{AshtekarKrishnan2004}.
That effect is dissipative, second order in the rate of change, and not of the
form of Eq.~(\ref{eq:F}), and it vanishes at the time-symmetric instant.

Finally, a remark that is speculative and forms no part of the argument. A
structure resembling our conclusion appears in the algebraic treatment of the
black-hole exterior, where the crossed product of the type III$_1$ algebra with
its modular automorphism group yields an algebra of type II$_\infty$, carrying a
trace and a well-defined entropy but admitting no factorization into subsystems
from which a patchwise entropy density would follow~\cite{Witten2022,CPW2023}.
Entropy without subsystems has the same shape as a total without a transport
rule. The settings differ and we derive neither statement from the other.

\section*{Data availability}

The data that support the findings of this article, and the code that
produces them, are openly available~\cite{CodeDeposit}.

\appendix

\section{Data, horizon and temperature}
\label{app:setup}

The Brill--Lindquist solution of the time-symmetric vacuum constraints is
conformally flat, $\gamma_{ij}=\psi^4\delta_{ij}$ with
$\psi=1+\sum_iG\mu_i/2\rho_i$. Each puncture is a separate asymptotically flat
end, and the mass measured in the $i$th end is
$m_i=\mu_i(1+\sum_{j\neq i}G\mu_j/2r_{ij})$. The lapse is
$\mathcal N=\chi/\psi$ with $\chi=2-\psi$ harmonic in the flat metric, and the
evolution vector is $\xi^a=\mathcal N\hat{u}^a$. At a moment of time symmetry
the horizon of each hole is the minimal surface surrounding its puncture,
$\rho_i=\varrho_i(\theta,\phi)$, axisymmetric here.

Each horizon element carries the Noether-charge temperature weight
$T=\kappa/2\pi$ with $\kappa=\xi_{a;b}\hat{N}^a\hat{u}^b$, which on the
slice is $\kappa=\partial_{\hat{N}}\mathcal N$. Because its normalization is tied
to $\xi^a$, it is the redshifted weight that enters the horizon charge and
Eq.~(\ref{eq:F}), not the proper Rindler temperature at a fixed-distance
stretched horizon. The contrast between this pointwise weight and an entropy
fixed only as a total is the subject of the paper.

Near the first hole the companion's contributions to $\psi$ and to $\chi$ are
$\pm G\mu_2/2\rho_2=\pm(G\mu_2/2\rt)(1+\rho\cos\theta/\rt+\cdots)$. The constant
term redshifts the hole's temperature by the companion's potential and gives
the monopole $(1-G\mu_2/\rt)/4G\mu_1$ at fixed $\mu_i$. The linear term is the
companion's uniform Newtonian field $G\mu_2/\rt^2$ at the hole and gives the
dipole $-3G\mu_2\cos\theta/16\rt^2$, the $\rho^2P_2$ tidal term entering only at
the next order. Re-expressed in the physical masses these give
Eq.~(\ref{eq:kappaexp}).

\section{What fixes the lapse}
\label{app:lapse}

Maximal slicing requires $\mathcal N=\chi/\psi$ with $\chi$ harmonic in the
flat metric. Restricted to functions with the same pole structure as $\psi$,
the admissible family is
\begin{equation}
\chi_{\vec\alpha}=1-\sum_i\alpha_i\frac{G\mu_i}{2\rho_i},
\end{equation}
and the small-sphere flux about the first puncture gives
\begin{equation}
Q_1=\frac{\mu_1}{4}\Bigl[(\alpha_1+1)+(\alpha_1-\alpha_2)\frac{G\mu_2}{2\rt}\Bigr].
\end{equation}
Requiring $2\sum_iQ_i=E_{\rm ADM}$ gives
$(\alpha_1-1)\mu_1+(\alpha_2-1)\mu_2=0$, which is one condition on two parameters
for a single configuration and selects $\alpha_1=\alpha_2=1$ only when imposed
identically in the masses. It then gives $\chi=2-\psi$.

This is not sufficient. Harmonic functions regular at infinity may carry
multipoles of any order about each puncture, and the Smarr flux is a monopole
integral that cannot see them. The term $c\cos\theta/\rho_1^2$ is harmonic,
leaves every $Q_i$ unchanged, and contributes $-16c\cos\theta/G^3\mu_1^3$ to the
normal derivative of $\chi$ at the horizon, hence a freely tunable dipole in
$\kappa$. Since the temperature anisotropy drives Eq.~(\ref{eq:F}), this freedom
would make the candidate arbitrary.

What excludes it is the inner end. Under the Kelvin transform about the first
puncture, with $r_0\equiv G\mu_1/2$, a harmonic $h$ maps to
$(r_0/\tilde\rho)h(r_0^2/\tilde\rho)$, and the lapse
$\mathcal N=\chi/\psi$ is invariant. A monopole $c/\rho$ maps to the constant
$c/r_0$, whereas the dipole maps to $c\,\tilde\rho\cos\theta/r_0^3$,
which grows linearly while $\tilde\psi\to1$, so the lapse diverges at the second
asymptotic end. Requiring a lapse of unit magnitude at every asymptotic end
therefore removes all multipoles above the monopole, and with the monopole
coefficients fixed as above gives $\chi=2-\psi$ uniquely. Equivalently,
$\chi/\psi\to-\alpha_i$ at each inner end, so the odd normalization
$\mathcal N\to-1$ there is $\alpha_i=1$ directly. The other unit-magnitude choice,
$\mathcal N\to+1$ ($\alpha_i=-1$), gives $\chi=\psi$, $\mathcal N\equiv1$ and
$\kappa\equiv0$, and violates the Smarr condition above.

The normalization is the standard one. For a single hole $\psi=2$ at the throat,
so $\mathcal N$ vanishes on the horizon and tends to $\pm1$ at the two ends,
which is the odd lapse of puncture slicings, the sign change expressing the
opposite time orientations of the two regions. Its normal derivative gives
$\kappa=1/4G\mu$, the Schwarzschild value for a Killing field normalized at
infinity, and exactly the value required by the isolated-horizon first law
$\delta E=\kappa\,\delta\A/8\pi G$ with $E=\mu$. For two holes, the monopole of
Eq.~(\ref{eq:kappaexp}) times $\A_1/8\pi G$ reproduces $\mu_1/2$ through the
order displayed, consistent with Eq.~(\ref{eq:Qexact}), which holds exactly.

\section{Horizon area}
\label{app:area}

Computing the induced metric on $\rho=\varrho_1(\theta)$ and integrating gives
Eq.~(\ref{eq:area}). Written in the Brill--Lindquist parameters it reads
\begin{equation}
\A_i=16\pi G^2\mu_i^2\Bigl(1+\frac{G\mu_j}{2\rt}\Bigr)^{\!2}
\Bigl(1-\frac{G^6\mu_i^4\mu_j^2}{1120\,\rt^{6}}+\cdots\Bigr),\qquad j\neq i.
\label{eq:vpform}
\end{equation}
Decomposing the area element in Legendre polynomials, its $P_0$ component is
unity through relative order $G^5$, the first nonzero monopole correction arising
at relative order $G^6$ from the quadratic quadrupolar response, and every higher
component integrates to zero against the measure, which is why the area responds
only to the monopole part of the companion's perturbation. Because the surface is
minimal, its own displacement contributes nothing at first order, so the leading
correction is second order in the induced quadrupole, whose relative amplitude is
$O(G^3)$.

Of the three determinations of the coefficient $-1/1120$ listed in
Sec.~\ref{sec:rigid}, the first two use the series; the variational solutions
for the minimal surface use none of it. We use two implementations of the
variational solver: that of Appendix~\ref{app:rebuild}, and a second one, with
$7$ or $10$ Legendre modes, for the tests reported here. The second differs from
the closed-form coefficient by $-8.0\times10^{-4}$, $-8.0\times10^{-5}$, and
$-8.0\times10^{-6}$ at $G=10^{-3}$, $10^{-4}$, and $10^{-5}$, with $G$ used here
as the formal expansion parameter and $\mu_1=1$, $\mu_2=2$, $\rt=10$ held fixed.
The difference is the next term of the series, whose relative size is
$-4G\mu_j/\rt$ when the correction is written at fixed $\mu_i$ as in
Eq.~(\ref{eq:vpform}), that is $-0.8\,G$ for the values just quoted, and
$-G(m_i+6m_j)/\rt$ when it is written in the physical masses as in
Eq.~(\ref{eq:area}). Explicitly, the bracket in Eq.~(\ref{eq:vpform}) continues
as $+G^7\mu_i^4\mu_j^3/280\,\rt^7$, and that in Eq.~(\ref{eq:area}) as
$+G^7m_i^4m_j^2(m_i+6m_j)/1120\,\rt^7$; the two forms differ because
$\mu_i=m_i-\tfrac12E_b$ with $E_b=O(G)$. Extrapolated in $1/\rt$ from
$\rt=20$, $40$, $80$, $160$, and $320$ at $G=1$, the coefficient from the second
implementation agrees with $-\mu_i^4\mu_j^2/1120$ to within $6\times10^{-6}$ for
$\mu_1{:}\mu_2=1{:}4$, $3\times10^{-7}$ for $1{:}2$, and $10^{-7}$ or better
for $1{:}1$, $2{:}1$, and $4{:}1$, identically with $7$ and $10$ Legendre
modes.

\subsection*{Comparison with earlier expansions}
\label{app:comparison}

Summing Eq.~(\ref{eq:vpform}) over both holes gives, through $O(\rt^{-6})$ and
with $G=1$,
\begin{equation}
16\pi\left[\mu_1^2+\mu_2^2+\frac{\mu_1\mu_2(\mu_1+\mu_2)}{\rt}
+\frac{\mu_1^2\mu_2^2}{4\rt^2}
\Bigl(2-\frac{\mu_1^4+\mu_2^4}{280\,\rt^4}\Bigr)\right],
\end{equation}
which is the expansion previously derived in Ref.~\cite{vanPutten2012}, with one
difference. That reference writes the last group, in its own notation, as
$2-M^8/280m^4a^4-m^8/280M^4a^4$ rather than $2-(\mu_1^4+\mu_2^4)/280\rt^4$. The
two agree for equal masses but differ by a factor $(\mu_1/\mu_2)^4$ otherwise,
and the printed form diverges as $\mu_2\to0$, where the perturbation of the first
hole vanishes and the correction must vanish with it; we take it to be a
misprint. The variational solution just described confirms the coefficient in
Eq.~(\ref{eq:vpform}) for unequal masses, independently of the expansions used
elsewhere.

\section{The relabeling identity}
\label{app:relabel}

Suppressing $\phi$ by axisymmetry and writing $dS=\tilde{s}(\theta,\rt)\,d\theta$,
any smooth orientation-preserving relabeling $\theta=g(\theta',\rt)$ fixing the
poles describes the same family of surfaces with the same physical measure,
$\tilde{s}'=\tilde{s}\,\partial_{\theta'}g$ and $T'=T$, and is characterized by its
label velocity $v=\partial_{\rt}g$. Expanding
$\partial_{\rt}[\tilde{s}\,\partial_{\theta'}g]$ by the product and chain rules and
comparing with $\partial_{\theta'}[T\tilde sv]$ yields the pointwise identity
\begin{equation}
T\,\partial_{\rt}\bigl[\tilde{s}\,\partial_{\theta'}g\bigr]
=T\,(\partial_{\rt}\tilde{s})\,\partial_{\theta'}g
+\partial_{\theta'}\bigl[T\tilde sv\bigr]
-(\partial_\theta T)\,v\,\tilde{s}\,\partial_{\theta'}g ,
\end{equation}
valid for arbitrary smooth $T$, $\tilde{s}$ and $g$. Integrating over $\theta'$,
the total derivative contributes only boundary terms, which vanish because
$\tilde{s}\propto\sin\theta$ and $v\to0$ at the poles, and this gives
Eq.~(\ref{eq:relabelaxi}), the axisymmetric case of Eq.~(\ref{eq:relabel}).

For a closed-form example take
$\tilde{s}=\tfrac12S_0\sin\theta[1+\alpha(\rt)\cos\theta]$ and
$T=T_0+T_1\cos\theta$, for which $\int\tilde{s}\,d\theta=S_0$ for all $\rt$. In the
fixed-$\theta$ labeling $F=S_0T_1\alpha'/3$, which is positive for the physical
signs $T_1<0$ and $\alpha'<0$. Relabeling by cumulative entropy makes the measure
uniform and separation independent, so $F'=0$ identically, and
Eq.~(\ref{eq:relabelaxi}) accounts for the difference exactly.

\section{The density freedom}
\label{app:density}

\subsection{The density under a fixed identification}

Suppose instead that the identification is held fixed, with horizon points
labeled by the angle subtended at the puncture, and that the density is
required to satisfy $\partial_{\rt}(dS)=0$ pointwise, so that the candidate
vanishes for every temperature profile. Together with the isolated-hole limit
this selects a density, consistently only through the order at which
Eq.~(\ref{eq:area}) makes the total separation independent, since
$\partial_{\rt}(dS)=0$ and $\oint dS=\A_1/4G$ cannot both hold once the total
depends on $\rt$. Write the area element as
$\dA_1=4G^2m_1^2\sin\theta\,\{1+\cdots\}\,d\theta\,d\phi$, where the brace
collects the separation-dependent terms and has $P_0$ component unity through
relative order $G^5$. The separation-independent part of $\dA_1/4G$ is
$Gm_1^2\sin\theta\,d\theta\,d\phi$, so $dS=Gm_1^2\sin\theta\,d\theta\,d\phi$, with
$\oint dS=\A_1/4G$ through relative order $G^5$. The two conditions do not by
themselves fix the angular density, since any separation-independent density
with the same integral would serve; what selects this one is the isolated-hole
limit $dS\to\dA_0/4G$ as $\rt\to\infty$, with
$\dA_0=4G^2m_1^2\sin\theta\,d\theta\,d\phi$ the area element of the isolated
throat. Under this labeling the density
departs from one quarter of the area element by
\begin{equation}
dS-\frac{\dA_1}{4G}
=-\frac{G^3m_1^3m_2}{4\,\rt^{2}}\,\sin2\theta\,d\theta\,d\phi+O(G^4),
\label{eq:domega}
\end{equation}
computed here through relative $O(G^5)$ and integrating to zero at each order,
because the brace has no monopole beyond unity. Effacement strictly requires
only that the force vanish, a single scalar condition, whereas
$\partial_{\rt}(dS)=0$ is pointwise and therefore stronger; the weaker
requirement leaves residual freedom, and this density is its natural solution
rather than its only one.

\subsection{The form of the required density}
\label{app:jkm}

At leading order Eq.~(\ref{eq:domega}) may be written
$dS-\dA_1/4G=-(R_h^3/6G)(\Delta\kappa)\,\dA$, with $R_h=2Gm_1$ and $\Delta$
the Laplacian on the horizon. This has the form of an exact local surface
differential of the kind relevant to the Jacobson--Kang--Myers ambiguity.
Whether it belongs to the admissible local and covariant class is not
established here, and in particular our $\kappa$ itself depends on the chosen
evolution vector. Whether the form persists beyond leading order we do not
determine. Nothing in the main text depends on it, Eq.~(\ref{eq:relabel}) being
independent of the form of the density.

\subsection{Why existing constructions do not settle the density}
\label{app:dynamical}

At a moment of time symmetry both null expansions of a minimal surface vanish.
The horizon therefore resembles a bifurcation surface, but it is not the
bifurcation surface of a Killing horizon, and the spacetime is not stationary.
That distinction is what matters. Iyer and Wald~\cite{IyerWald1994} exempt the
Jacobson--Kang--Myers freedom only for the entropy on a cross-section of a
stationary black hole and for the first variation at the bifurcation surface of
one, and state that the freedom matters for a dynamical black hole, more
severely on an arbitrary cross-section of a nonstationary one. Neither exemption
applies here.

Nor do the later constructions fix it. Visser and Yan~\cite{VisserYan2024} show
the ambiguities to be exact differentials that vanish on integration over a
compact cross-section and then set them to zero without loss of generality,
which holds for the integral only; Wall and Yan~\cite{WallYan2024} argue
in the same way. Their results, and those of
Refs.~\cite{HollandsWaldZhang2024,Wall2015,Dong2014}, are established at first
order in perturbations of a stationary black hole, whereas the candidate of
Eq.~(\ref{eq:F}) is second order in the companion's field.
Reference~\cite{HollandsWaldZhang2024} notes that for source-free first-order
perturbations the entropy change vanishes and the leading change appears at
second order, which is Eq.~(\ref{eq:area}) seen from the other side; and at
second order it identifies as unresolved the residual rescaling, in its
notation, $V\to f(x^A)V$ of the affine parameter of the generators by an
arbitrary function on the cross-section, together with the resulting need to
fix its time-translation field off the horizon when the metric has no horizon
Killing field. Finally, the
dynamical-horizon segments of Refs.~\cite{Ashtekar2026,Ashtekar2026b} require
the expansion of the second null normal to be nowhere zero, which fails
identically here.

\section{The force series}
\label{app:force}

Inserting the area element and the nonuniform temperature into
Eq.~(\ref{eq:F}) and summing over the two horizons gives
\begin{align}
F={}&\frac{G^4m_1^2m_2^2(m_1+m_2)}{8\,\rt^5}\nonumber\\
&-\frac{G^5m_1^2m_2^2\bigl[5(m_1^2+m_2^2)+34m_1m_2\bigr]}{32\,\rt^6}
\nonumber\\
&+\frac{3G^6m_1^2m_2^2\bigl[27(m_1^3+m_2^3)
+343m_1m_2(m_1+m_2)\bigr]}{448\,\rt^7}\nonumber\\
&-\frac{G^7m_1^2m_2^2}{1120\,\rt^8}\bigl[175(m_1^4+m_2^4)
\nonumber\\
&\qquad+4294m_1m_2(m_1^2+m_2^2)+7875m_1^2m_2^2\bigr] +\cdots .
\label{eq:forceseries}
\end{align}
Every coefficient displayed is an exact rational. The minimal-surface condition
for $\varrho_1$ is solved order by order in $1/\rt$ in a Legendre basis, each
order being uniquely determined because the linearized operator about the
Schwarzschild throat has no kernel, and the area element and $\kappa$ are then
evaluated on the resulting shape. The series has been carried to
$O(\rt^{-10})$; the terms beyond those displayed are not needed here. The leading
term gives Eq.~(\ref{eq:ratio}), and it is the $O(G^6)$ term that fixes the
coefficient in Eq.~(\ref{eq:area}), since reproducing it requires the monopole of
the area element two orders beyond the leading expansion.

\subsection*{Invariant and identification-dependent pieces}
\label{app:pieces}

Split the integrand exactly as
$\int T\,\partial_{\rt}(dS)=\partial_{\rt}\oint T\,dS-\int(\partial_{\rt}T)\,dS$.
Differentiating $Q_i=\mu_i/2$ at fixed physical masses, with
$\mu_i=m_i-\tfrac12E_b$ and $E_b$ the closed form of Eq.~(\ref{eq:Pexact}),
\begin{equation}
\frac{dQ_i}{d\rt}=\frac{E_b}{2\sqrt{(GM+2\rt)^2-4G^2m_1m_2}},
\label{eq:dQdr}
\end{equation}
which expands as $Gm_1m_2/4\rt^2-G^2m_1m_2M/4\rt^3+\cdots$ and is the same for
both holes, so that $\sum_idQ_i/d\rt=\tfrac12\,dE_{\rm ADM}/d\rt$ at every order,
as the Smarr relation requires. The identification-dependent remainder
$\int(\partial_{\rt}T)\,dS=dQ/d\rt-\int T\,\partial_{\rt}(dS)$ coincides with
Eq.~(\ref{eq:dQdr}) term by term through $O(G^3)$, and the two first differ at
$O(G^4)$, where
$-\tfrac18 G^4m_1m_2(m_1^3+7m_1^2m_2+6m_1m_2^2+m_2^3)\rt^{-5}$ replaces
$-\tfrac18 G^4m_1m_2(m_1+m_2)(m_1^2+5m_1m_2+m_2^2)\rt^{-5}$. The difference is
the single-horizon candidate $G^4m_1^3m_2^2/8\rt^5$.

\section{Post-Newtonian size of the candidate}
\label{app:pn}

In the reduced variables of Ref.~\cite{DJS2000inv}, with
$\hat{\boldsymbol n}=\hat{\boldsymbol r}/\hat{r}$, the two 3PN ambiguity terms may
be written
\begin{equation}
\delta\hat{H}_{\rm amb}=
\frac{\nu\,\omega_{\rm static}}{\hat{r}^4}
+\frac{\nu^2\omega_{\rm kinetic}}{\hat{r}^3}
\bigl[\hat{p}^2-3(\hat{\boldsymbol n}\!\cdot\!\hat{\boldsymbol p})^2\bigr].
\label{eq:Hamb}
\end{equation}
To this order an infinitesimal canonical transformation generated by
$g=\lambda\,(\hat{\boldsymbol r}\!\cdot\!\hat{\boldsymbol p})/\hat{r}^3$ changes the
Hamiltonian, with the convention $\hat{H}\to\hat{H}+\{g,\hat{H}_{\rm N}\}$ and
$\hat{H}_{\rm N}=\hat{p}^2/2-1/\hat{r}$, by
\begin{equation}
\{g,\hat{H}_{\rm N}\}=\lambda\left\{
\frac{\hat{p}^2-3(\hat{\boldsymbol n}\!\cdot\!\hat{\boldsymbol p})^2}{\hat{r}^3}
-\frac{1}{\hat{r}^4}\right\}.
\label{eq:canonshift}
\end{equation}
Choosing $\lambda=-\nu^2\omega_{\rm kinetic}$ removes the kinetic ambiguity and
leaves the static coefficient
\begin{equation}
\sigma(\nu)=\nu\,\omega_{\rm static}+\nu^2\omega_{\rm kinetic}.
\label{eq:sigma}
\end{equation}
The unit relative coefficient thus follows directly from a canonical
transformation, in agreement with the invariant combination identified in
Refs.~\cite{DJS2000inv,DJS2000poincare}.

For a residual $\delta\hat{H}=\sigma/\hat{r}^4$ added to the Newtonian Hamiltonian,
circular orbits satisfy $J^2=\hat{r}-4\sigma/\hat{r}^2$ and
$\hat{E}=-1/2\hat{r}-\sigma/\hat{r}^4$, with $J$ the reduced angular momentum,
whence, with $x=(GM\Omega)^{2/3}$ and $\Omega$ the orbital frequency,
\begin{equation}
\hat{E}(x)=-\frac{x}{2}\Bigl(1+\frac{10\sigma}{3}x^3\Bigr)
=-\frac{x}{2}\Bigl(1+\frac{5\nu}{48}x^3\Bigr)
\label{eq:Ex}
\end{equation}
for $\sigma=\nu/32$. Writing the binding energy as
$\hat{E}=-\tfrac{x}{2}(1+\sum_ke_kx^k)$, this is to be compared with the third
post-Newtonian coefficient of general relativity,
$e_3=-\tfrac{675}{64}+(\tfrac{34445}{576}-\tfrac{205}{96}\pi^2)\nu
-\tfrac{155}{96}\nu^2-\tfrac{35}{5184}\nu^3$, which is $-10.55$ in the test-mass
limit but only $-0.967$ at $\nu=\tfrac14$, so the candidate is $2.7\%$ of the
third post-Newtonian coefficient for equal masses.

The candidate shifts the third post-Newtonian phasing coefficient by
$\delta\varphi_6=-35\nu/3$. The conservative modification supplies
$-80\,\delta e_3=-25\nu/3$ of this; the remainder is not optional. Write the
flux as $\mathcal F=\tfrac{32}{5}\nu^2x^5(1+\sum_kf_kx^k)$. Because
$\delta\hat{H}=\sigma/\hat{r}^4$ moves $\hat{r}$ at fixed $\Omega$ by the relative
amount $-\tfrac43\sigma x^3$, the Newtonian quadrupole flux, which goes as
$\hat{r}^4\Omega^6$, shifts by $\delta f_3=-16\sigma/3=-\nu/6$, and
$\partial\varphi_6/\partial f_3=+20$ adds a further $-10\nu/3$. The total is
$2.6\times10^{-3}$ of the general-relativistic value at $\nu=\tfrac14$, falling to
$1.9\times10^{-3}$ at $\nu=0.1$. Here $\varphi_6$ is the nonlogarithmic third
post-Newtonian coefficient, which carries the constant $-(6848/21)\ln4$, since
the parametrization of Ref.~\cite{GWTC4TGR} assigns to $\varphi_{6l}$ only what
is logarithmic in frequency.

The sensitivities follow from the stationary-phase construction. Writing
$dt/dv=-E'(v)/\mathcal F(v)=(5M/32\nu v^9)\,[1+\sum_kd_kv^{2k}]$, with
$v=x^{1/2}$ the post-Newtonian velocity parameter (not the label velocity),
integrating for $t(v)$ and $\phi(v)$, and assembling $\Psi=2\pi ft-2\phi-\pi/4$
gives $\varphi_{2k}=\tfrac{1}{3}d_k[40/(5-2k)-40/(8-2k)]$, whose $k=1$ case
returns $\varphi_2=\tfrac{3715}{756}+\tfrac{55}{9}\nu$. At $k=3$ the weight is
$-20$; $e_3$ enters $d_3$ only through the term $4e_3v^6$ of $E'$, and $f_3$
through $-f_3$, so $\partial\varphi_6/\partial e_3=-80$ and
$\partial\varphi_6/\partial f_3=+20$. Both figures are ratios of coefficients of
the same power $x^3$ and involve no choice of separation; they are meaningful
wherever the post-Newtonian expansion is, that is for $\rt\gg GM$, and say nothing
about the near zone, where the $1/\rt$ series fails and, in Brill--Lindquist data,
a common apparent horizon forms at separations of order $GM$. The absolute size
of the candidate at a given separation is Eq.~(\ref{eq:ratio}): for equal masses,
$3\times10^{-5}$ of the Newtonian attraction at $\rt=10\,GM$ and
$3\times10^{-8}$ at $100\,GM$. Since the next term of the force series is
$-\tfrac{11}{4}GM/\rt$ relative to the leading one for equal masses, the leading
term alone is accurate to ten percent only for $\rt\gtrsim30\,GM$.

\section{The static transverse-traceless field}
\label{app:tt}

By the physical slice we mean black-hole data on the slice where the momenta
vanish whose field for $Gm_i\ll\rho_i\ll\rt$ is the post-Newtonian one. It
differs from Brill--Lindquist data by a transverse-traceless (TT) part and by
the changes that part induces, at higher order, in the conformal factor and the
lapse. This appendix shows that the difference leaves the candidate unchanged
through third post-Newtonian order, as stated in Secs.~\ref{sec:corroboration}
and~\ref{sec:isometry}.
In ADMTT gauge $\gamma_{ij}=\psi^4\delta_{ij}+h^{\rm TT}_{ij}$, and with the
momenta zero the leading TT field obeys [Eqs.~(26) and (27) of
Ref.~\onlinecite{JS1998}, with $\phi_{(2)}=8(\psi-1)$]
\begin{equation}
\nabla^2h^{\rm TT}_{ij}=-16\,\delta^{{\rm TT}\,kl}_{ij}\,
\partial_k\psi\,\partial_l\psi,
\qquad
\psi=1+\sum_i\frac{Gm_i}{2\rho_i},
\label{eq:TTfield}
\end{equation}
with $\nabla^2$ the flat Laplacian,
$\delta^{{\rm TT}\,kl}_{ij}=\tfrac12(\Pi_{ik}\Pi_{jl}+\Pi_{il}\Pi_{jk}
-\Pi_{ij}\Pi_{kl})$ and $\Pi_{ij}=\delta_{ij}-\partial_i\partial_j\nabla^{-2}$;
to this order $\mu_i$ may be replaced by $m_i$. The field is of relative order
$(GM/\rt)^2$. At the instant the momenta vanish the TT momentum also vanishes at
this order, so the slice is time symmetric through the order used here: as for
Brill--Lindquist data, the horizons are minimal surfaces and the maximal-slicing
lapse is harmonic in the physical three-metric.

Three identities do the work. With $w_i\equiv1/\rho_i$,
\begin{align}
&\partial_kw_i\,\partial_lw_i=\tfrac18\,\partial_k\partial_l(w_i^2)
+\tfrac14\,\delta_{kl}\,w_i^4,
\label{eq:TTself}\\
&\partial_kw_1\,\partial_lw_2+\partial_lw_1\,\partial_kw_2\nonumber\\
&\qquad=\partial_k\partial_l(w_1w_2)-w_1\,\partial_k\partial_lw_2
-w_2\,\partial_k\partial_lw_1,
\label{eq:TTcross}\\
&\varpi\,\partial_k\partial_lw=\tfrac12(\partial_kU_l+\partial_lU_k),
\qquad U_l=\varpi\,\partial_lw-\varpi_l\,w,
\label{eq:TTaffine}
\end{align}
the last holding for any $w$ and any affine function
$\varpi=\varpi_0+\varpi_px^p$, with $\varpi_l\equiv\partial_l\varpi$. The
projector annihilates double gradients, pure traces and symmetrized gradients.
By Eq.~(\ref{eq:TTself}) the self terms in Eq.~(\ref{eq:TTfield}) source
nothing away from the punctures, as the conformal flatness of a single isotropic
Schwarzschild hole requires; near the punctures the field is that of the matched
black-hole solution. Write $w_2=w_2^{\rm aff}+w_2^{\rm rest}$, where
$w_2^{\rm aff}$ is the affine Taylor polynomial of $w_2$ at the first puncture,
carrying the companion's potential and uniform field there, and split $w_1$ in
the same way about the second puncture. Equations~(\ref{eq:TTcross}) and
(\ref{eq:TTaffine}) then give, exactly,
\begin{equation}
\nabla^2h^{\rm TT}_{ij}=4G^2m_1m_2\,\delta^{{\rm TT}\,kl}_{ij}
\bigl(w_2^{\rm rest}\,\partial_k\partial_lw_1
+w_1^{\rm rest}\,\partial_k\partial_lw_2\bigr),
\label{eq:TTrest}
\end{equation}
since the terms removed are a double gradient and symmetrized gradients of
vector fields that vanish at infinity. The companion's potential and uniform
field therefore source no TT field. In the Brill--Lindquist data the same
uniform field is pure gauge, generated by the special conformal vector of
Eq.~(\ref{eq:sct}); here it drops out of the TT sector identically.

Near the first puncture use Cartesian coordinates $x^p$ centered on it, with
$\rho=|x|$, $\hat{x}=x/\rho$ and $\hat{z}$ the unit vector toward the companion.
Then $w_2^{\rm rest}=\tfrac12w_{2,pq}x^px^q+O(\rho^3)$, with
$w_{2,pq}\equiv\partial_p\partial_qw_2(0)=(3\hat{z}_p\hat{z}_q-\delta_{pq})/\rt^3$
the companion's traceless tidal tensor ($-\mathcal E_{pq}/Gm_2$ in the notation
of Sec.~\ref{sec:pnhorizon}), and the bracket in Eq.~(\ref{eq:TTrest}) is
\begin{equation}
\frac1\rho\Bigl[\tfrac12\,w_{2,pq}\hat{x}^p\hat{x}^q\,
(3\hat{x}_k\hat{x}_l-\delta_{kl})+w_{2,kl}\Bigr]+O(\rho^0),
\label{eq:TTsource}
\end{equation}
even under $x\to-x$ and homogeneous of degree $-1$. The kernel of
$\nabla^{-2}\delta^{\rm TT}$ is also even and homogeneous of degree $-1$, its
Fourier symbol being $-k^{-2}$ times a projector even in $k$, so this part of
the source produces a field even about the puncture, of the form constant
$+\,\rho\,\mathcal K_{ij}(\hat{x})$; a term $\rho\ln\rho$ would need a source
component resonant with the harmonic polynomials of degree one, which are odd.
The rest of the source adds a constant, terms linear in $x$, and terms of order
$\rho^2$. Hence, for $Gm_1\ll\rho\ll\rt$,
\begin{equation}
h^{\rm TT}_{ij}=\mathcal C_{ij}+\mathcal D_{ijp}\,x^p
+\rho\,\mathcal K_{ij}(\hat{x})+O(\rho^2),
\label{eq:TTlocal}
\end{equation}
with $\mathcal K$ even and of order $G^2m_1m_2/\rt^3$. No term of order
$\rho^0$ is odd under $x\to-x$: it could only come from an odd source of degree
$-2$, which is the uniform-field term removed above. Expanding the closed-form
solution of Eq.~(\ref{eq:TTfield}) [the static part of Eq.~(A20) of
Ref.~\onlinecite{JS1998}] confirms Eq.~(\ref{eq:TTlocal}) and gives
\begin{equation}
\begin{aligned}
\mathcal C_{ij}&=\frac{3G^2m_1m_2}{2\rt^2}\,(\delta_{ij}-3\hat{z}_i\hat{z}_j),\\
\mathcal D_{ijp}&=\frac{G^2m_1m_2}{\rt^3}\,(\delta_{ij}\hat{z}_p+\delta_{ip}\hat{z}_j
+\delta_{jp}\hat{z}_i-5\hat{z}_i\hat{z}_j\hat{z}_p).
\end{aligned}
\label{eq:TTCD}
\end{equation}
Continued to the horizon, $\rho\simeq Gm_1/2$, the three terms of
Eq.~(\ref{eq:TTlocal}) are of relative order $(GM/\rt)^2$, $(GM/\rt)^3$ and
$(GM/\rt)^3$.

In the overlap region $Gm_1\ll\rho\ll\rt$ the terms
$\mathcal C_{ij}+\mathcal D_{ijp}x^p$ are the field of sources away from the
hole. There a constant metric perturbation is removed by a linear change of
coordinates and a linear one by a quadratic change, so, in the
matched-asymptotic construction of
Sec.~\ref{sec:pnhorizon}~\cite{TaylorPoisson2008}, $\mathcal C$ and
$\mathcal D$ alter only the coordinate labels of the horizon, not its geometry.
The feedback of $\mathcal C$ into the strong field is consistent with this. The
Hamiltonian constraint gives
$\nabla^2\psi\supset\tfrac12h^{\rm TT}_{ij}\partial_i\partial_j\psi$ [Eqs.~(17)
and (A7) of Ref.~\onlinecite{JS1998}] and maximal slicing gives
$\nabla^2\delta\mathcal N\simeq h^{\rm TT}_{ij}\partial_i\partial_j\mathcal N$,
so that $h^{\rm TT}=\mathcal C$ changes $\psi^4$ by
$-(Gm_1/2\rho)\,\hat{x}\!\cdot\!\mathcal C\!\cdot\!\hat{x}$ and the lapse by
$(Gm_1/2\rho)\,\hat{x}\!\cdot\!\mathcal C\!\cdot\!\hat{x}$. These are exactly the
trace part, in ADMTT form, and the lapse change of the coordinate change
generated by $\tfrac12\mathcal C_{ij}x^j$. The hole responds physically to the
external part only through its curvature, a correction of relative order
$Gm_1/\rt$ to the Newtonian tide.

The even term $\rho\,\mathcal K$ is different. It is generated near the hole by
the hole's own field acting with the companion's tide, and it is the part of
the hole's static tidal response that conformally flat data cannot represent:
Brill--Lindquist data have vanishing Cotton--York tensor, whereas
$\rho\,\mathcal K$ gives the physical slice one of order
$G^2m_1m_2/(\rt^3\rho^2)$ near the hole, to which $\mathcal C$ and $\mathcal D$
do not contribute. It is therefore a genuine difference of horizon geometry, as
large on the horizon as the tide's own effect and, being even and linear in the
traceless tidal tensor, confined to $\ell=2$ at this order. Since the horizon is a
minimal surface, a perturbation changes its area at first order only through
its monopole on the horizon, which vanishes for $\rho\,\mathcal K$ as for any
perturbation linear in the traceless tidal tensor; the area at fixed mass
therefore changes first at relative order $(GM/\rt)^6$, quadratic in the tide.

The candidate uses puncture labels, which are coordinate dependent; only $s_0$,
the total entropy divided by $4\pi$, is unchanged by every relabeling. At
relative order $(GM/\rt)^2$ only $\mathcal C$ acts, and any effect linear in a
constant traceless tensor on the spherical throat has $\ell=2$: an $\ell=0$ part would
be proportional to $\mathcal C_{ii}=0$, and odd $\ell$ cannot be formed. At that
order, therefore, $\delta s_0=\delta s_1=0$ and
$\delta s_2=O[(GM/\rt)^2]\,s_0$, while $T$, a scalar, has its existing
anisotropy moved but no new anisotropy created: $\mathcal C$ mixes $T_1$ into
$\ell=1,3$ at $O[(GM/\rt)^4]\,T_0$. Odd relabelings first arise from $\mathcal D$,
at relative order $(G^2m_1m_2/\rt^3)(Gm_1)$, so that
$\delta s_1=O[(GM/\rt)^3]\,s_0$, while they move the temperature dipole only at
$O[(GM/\rt)^5]\,T_0$. The term $\rho\,\mathcal K$ changes $s_2$ and $T_2$ at
$O[(GM/\rt)^3]$. The field also changes the lapse. Near the hole this is the
image of $\mathcal C$ and $\mathcal D$ under the coordinate changes above; the
part sourced away from the hole changes the redshift there, by
$\delta T_0/T_0=O[(GM/\rt)^3]$ and $\delta T_1/T_0=O[(GM/\rt)^4]$, which is not a
relabeling.

In Eq.~(\ref{eq:forceLmain}), with $T_1/T_0$ and $s_1/s_0$ of order
$(GM/\rt)^2$ and $T_2/T_0$ and $s_2/s_0$ of order $(GM/\rt)^3$, these give
$\delta F_{\ell=1}/F_{\rm N}=O[(GM/\rt)^4]$ from $T_1\,\partial_{\rt}\delta s_1$,
$\delta F_{\ell=2}/F_{\rm N}=O[(GM/\rt)^4]$ from $T_2\,\partial_{\rt}\delta s_2$,
and $\delta F_{\ell=0}$ only beyond fourth post-Newtonian order; the lapse change
first enters at fifth order. In the language of Eq.~(\ref{eq:relabel}), the
label velocity of $\mathcal C$ is of the same order as the M\"obius velocity of
Appendix~\ref{app:moebius}, which carries the whole 3PN candidate, but it is
proportional to $\partial_\theta P_2$ and so pairs at leading order only with
$T_2$. The candidate on the physical slice therefore equals its
Brill--Lindquist value through third post-Newtonian order, in every multipole,
and $\delta\omega_{\rm static}=1/32$ is common to both. The two can differ at fourth
order, through $\mathcal D$ at $\ell=1$ and through $\mathcal C$ at $\ell=2$, both of
them relabelings. The first statement of Eq.~(\ref{eq:l1hiermain}) therefore
refers to Brill--Lindquist data in puncture labels, while the second holds on
both slices as a statement of order, since $\mathcal C$ and $\mathcal D$ are not
horizon geometry and $\rho\,\mathcal K$ and the lapse change first enter at
fifth order. In balanced labels $\mathcal C$ and $\mathcal D$ drop out, so the
balanced-frame force vanishes through fourth order on both slices. Its
fifth-order coefficient in Eq.~(\ref{eq:Fbalmain}), like that in
Eq.~(\ref{eq:F5PNmain}), is a Brill--Lindquist value, which $\rho\,\mathcal K$
can change on the physical slice, though not its order.

\section{The special conformal map}
\label{app:isometry}

For $\zeta^i=\beta(2z\,x^i-\rho^2\delta^i_z)$ one finds
$\partial_i\zeta_j+\partial_j\zeta_i=4\beta z\,\delta_{ij}$ by direct
computation. Writing $\psi_0=a+G\mu_1/2\rho$,
\begin{equation}
\Lie_\zeta(\psi_0^4\delta_{ij})
=4\psi_0^3\bigl(\zeta\!\cdot\!\nabla\psi_0+\beta z\,\psi_0\bigr)\delta_{ij} ,
\end{equation}
and since $\zeta\cdot\nabla\psi_0=-\beta z\,G\mu_1/2\rho$ the terms in $\mu_1$
cancel, leaving $4\psi_0^3\,a\beta z\,\delta_{ij}$. Choosing $\beta=c_1/a$ makes
this the perturbation $4\psi_0^3c_1z\,\delta_{ij}$ generated by the linear term of
Eq.~(\ref{eq:nearfield}).

The response of the horizon follows from the linearized minimal-surface
condition. In the conformally flat metric the mean curvature of a surface is
$\psi^{-2}(H_{\rm f}+4\partial_{\hat{N}_{\rm f}}\ln\psi)$, with $H_{\rm f}$ its
flat mean curvature, so the throat of $\psi_0$ sits at $\rho_0=G\mu_1/2a$, where
$\psi_0=2a$. Perturb $\psi_0$ by $q_\ell\rho^\ell P_\ell(\cos\theta)$ and the
surface by $\delta\ln\varrho_1=h_\ell P_\ell(\cos\theta)$. Since
$H_{\rm f}=[2-2h_\ell P_\ell+\ell(\ell+1)h_\ell P_\ell]/\rho_0$ to first order
and $\partial_\rho\ln\psi_0$ has derivative $3/4\rho_0^2$ at the throat, the
condition becomes
$(\ell^2+\ell+1)h_\ell/\rho_0+(2\ell+1)q_\ell\rho_0^{\ell-1}/a=0$, so that
\begin{equation}
h_\ell=-\frac{2\ell+1}{\ell^2+\ell+1}\,\frac{q_\ell\rho_0^\ell}{a}.
\label{eq:hl}
\end{equation}
For the uniform field, $q_1=c_1$, this gives
$\varrho_1=(G\mu_1/2a)[1-(G\mu_1c_1/2a^2)\cos\theta]$, and the dipole of the area
element in puncture labels is $+(G^2m_1m_2/2\rt^2)\cos\theta$, identically the
Jacobian of the M\"obius map induced on the horizon by $\zeta$. The dipole of
$\kappa$ is $-3G\mu_2\cos\theta/16\rt^2$, in agreement with
Eq.~(\ref{eq:kappaexp}). For the tidal quadrupole, $q_2=G\mu_2/2\rt^3$ at leading
order, it gives $h_2=-\tfrac{10}{7}\eta$ with $\eta=G^3m_1^2m_2/16\rt^3$, the
displacement quoted in Sec.~\ref{sec:isometry}.

\subsection*{The leading dipole as a M\"obius reparametrization}
\label{app:moebius}

Write $\varrho\equiv\varrho_1(\theta)$ for the horizon locus and
$\varrho'=\partial_\theta\varrho$. The Gaussian curvature $K$ of the induced
metric
$\psi^4[(\varrho^2+\varrho'^2)\,d\theta^2+\varrho^2\sin^2\theta\,d\phi^2]$,
expanded
in Legendre components with $\oint K\,\dA=4\pi$ at every order, is a pure
monopole through relative order $G^2$, where it equals $1/(2Gm_1)^2$ exactly; its
first non-monopole component is a $P_2$ term at relative order $G^3$. A metric of
constant curvature on $S^2$ is round, so through the order that produces
Eq.~(\ref{eq:ratio}) the horizon is isometric to the round sphere of areal radius
$2Gm_1$, independent of $\rt$.

The $O(G^4)$ candidate is then traced to a single product: the $P_1$ component of
$\dA_1$ at relative order $G^2$, namely
$4G^2m_1^2\sin\theta\,(G^2m_1m_2/2\rt^2)\cos\theta\,d\theta\,d\phi$, against the
$P_1$ component of $T$ at the same order, $-(3Gm_2/32\pi\rt^2)\cos\theta$. The
higher monopoles of $\dA_1$ vanish, and the leading correction to $T$ is itself a
monopole. An $\ell=1$ change of the area density on a round sphere is the Lie
derivative of the round area form along a conformal Killing field. Writing
$\dA_0$ for the area form of that round sphere and $\nabla_0$ for its gradient,
the field $X=R_h^2\,\nabla_0\cos\theta=-\sin\theta\,\partial_\theta$ satisfies
$\Lie_X\dA_0=-2\cos\theta\,\dA_0$, so that
\begin{equation}
\partial_{\rt}(\dA_1)\big|_{G^2}=\frac{G^2m_1m_2}{2\rt^3}\,\Lie_X\dA_0 .
\label{eq:liew}
\end{equation}
In the language of Eq.~(\ref{eq:relabelaxi}), the label velocity
$v=(G^2m_1m_2/2\rt^3)\sin\theta$ makes the density $\rt$-independent at this
order, and $\int(\partial_\theta T)\,v\,dS=G^4m_1^3m_2^2/8\rt^5$ is the whole
single-horizon candidate. The leading 3PN candidate is therefore the
temperature-weighted rate of a M\"obius reparametrization of the round reference
geometry, equivalently the drift of its center relative to the puncture.

The cumulative-entropy labeling of Appendix~\ref{app:relabel} is the axisymmetric
case of Moser's theorem~\cite{Moser1965}. Since $\A_1$ is $\rt$-independent
through relative order $G^5$, an identification under which $\dA_1$ is pointwise
constant exists with no symmetry assumption, and for it Eq.~(\ref{eq:F}) returns
zero for every temperature profile. Beyond relative order $G^5$, where $\A_1$ does
depend on $\rt$, the proportional identification, in which the labels are chosen
so that $dS=(\A_1/4G)\sin\theta\,d\theta\,d\phi/4\pi$ at every separation,
returns for the first hole the contribution $(Q_1/S_1)\,dS_1/d\rt$ exactly, with
$S_1=\A_1/4G$, since
$\int T\sin\theta\,d\theta\,d\phi/4\pi=Q_1/S_1$ in those labels. This is the
equilibrium force at the entropy-weighted mean temperature, the 5PN residual of
Eq.~(\ref{eq:area}), which the label velocity of Sec.~\ref{sec:corroboration}
removes in turn.

\section{Independent rebuilds}
\label{app:rebuild}

Every coefficient above has been checked against a rebuild that uses none of the
series. The horizon of the first hole is obtained by solving the minimal-surface
condition directly from the Brill--Lindquist data, by Newton iteration on the area
functional in a Legendre basis at sixty-digit precision, and the area, the
surface gravity, the Komar flux, and the force are then evaluated from that shape
alone. Orders are extracted by Richardson extrapolation in $1/\rt$ from
$\rt=100$, $200$, $400$, and $800$ ($\mu_1=1$, $\mu_2=2$, $G=1$), and every value
quoted below is the same with $9$ Legendre modes and $120$ quadrature nodes as
with $12$ and $160$.

The Komar flux is exact rather than approximate: $\oint\kappa\,\dA/4\pi G\mu_1-1$
is of order $10^{-61}$, the working precision, at every separation tested, as the
flat Gauss law of Sec.~\ref{sec:smarr} requires. The relative area defect times
$\rt^6$ extrapolates to $-0.003571428$ against the closed form
$-\mu_1^4\mu_2^2/1120=-0.003571429$, and the dipole of $\kappa$ times $\rt^2$ to
$-0.374999998$ against $-3G\mu_2/16$, confirming both the magnitude and the sign of
the temperature anisotropy. The single-horizon candidate, computed from the
rebuilt shapes at fixed physical masses $m_1=1$, $m_2=2$ by central differences
in $\rt$, is $0.9081$, $0.9527$, and $0.9760$ of $G^4m_1^3m_2^2/8\rt^5$ at
$\rt=100$, $200$, and $400$, and Richardson extrapolation gives $0.99997$.

A second, fully analytic rebuild solves the mean-curvature condition for the
horizon order by order in $1/\rt$ in exact rational arithmetic, in the variables
$\mu_2/\mu_1$ and $\mu_1/(2\rt+\mu_2)$. Carried to tenth relative order it
reproduces exactly every displayed coefficient of Eq.~(\ref{eq:forceseries}),
Eq.~(\ref{eq:area}) together with both forms of its $O(G^7)$ continuation in
Appendix~\ref{app:area}, Eq.~(\ref{eq:kappaexp}), the coefficients of
Appendix~\ref{app:pieces}, and the multipole decomposition of
Sec.~\ref{sec:isometry}; extended to Eq.~(\ref{eq:Beqmain}) it gives
Eq.~(\ref{eq:mottclosed}). It agrees with the variational rebuild to all digits
compared. The multipole decomposition has been checked independently by a
spectral solver at 35 to 45 digits, with Legendre truncations of 16, 22, and 28
giving the same non-dipole fractions to twelve digits.

The rebuild also exhibits the fixed-$\mu_i$ path directly. Differentiating at
fixed $\mu_i$ rather than at fixed physical masses returns
$-G\mu_1\mu_2/2\rt^2$ for the first horizon at leading order, and through
$O(G^4)$ $-G\mu_1^2\mu_2/2m_1\rt^2+G^4\mu_1^3\mu_2^2/8\rt^5+O(G^5)$: minus one half
of the Newtonian attraction per horizon, opposite in sign to the candidate and
larger than it by $4\rt^3/G^3m_1^2m_2$, more than seven orders of magnitude at
$\rt=100\,GM$. Summed over the two horizons this reproduces the Newtonian-scale
term of Ref.~\cite{vanPutten2012}. It is not an additional interaction: at fixed
$\mu_i$ one is moving on a constant-$E_{\rm ADM}$ path, so the entropy derivative
re-expresses the ordinary energy gradient, as discussed in Sec.~\ref{sec:path}.
Its algebraic origin is simple. At fixed $\mu_i$ the invariant piece
$\partial_{\rt}\oint T\,dS$ vanishes identically, since $Q_i=\mu_i/2$, so the
entire fixed-$\mu_i$ entropy term is $-\int(\partial_{\rt}T)|_{\mu}\,dS$, the
derivative of the companion's redshift $(1-G\mu_2/\rt)/4G\mu_1$ of the mean
surface gravity.

\section{The tube equation}
\label{app:tube}

Let the MOTS move within the slices with normal speed $f$ while the slices
advance with lapse $\mathcal N$ and zero shift, so that the tube is generated by
Eq.~(\ref{eq:ABmain}). At $t=0$ the extrinsic curvature vanishes and so does the
mean curvature $q^{ij}D_i\hat{N}_j$ of the MOTS. To first order in $t$ the three-metric
is unchanged, since $\partial_t\gamma_{ij}=-2\mathcal NK_{ij}$, while the vacuum
evolution equation gives $\partial_tK_{ij}=-D_iD_j\mathcal N+\mathcal NR_{ij}$,
with $R_{ij}$ the Ricci tensor of the slice. Hence
\begin{equation}
\partial_t\theta_{(k)}=\delta_f\bigl(q^{ij}D_i\hat{N}_j\bigr)-q^{ij}\partial_tK_{ij},
\end{equation}
where the variation of the mean curvature under a normal displacement $f$ in a
fixed three-metric is
$\delta_f(q^{ij}D_i\hat{N}_j)=-\Delta f-(|b|^2+R_{\hat{N}\hat{N}})f$, with
$b_{AB}$ the extrinsic curvature of the MOTS in the slice and $\Delta$ its
Laplacian. On a minimal surface $q^{ij}D_iD_j\mathcal N=\Delta\mathcal N$, and the
Hamiltonian constraint at time symmetry gives $R=0$, so that
$q^{ij}R_{ij}=-R_{\hat{N}\hat{N}}$ and
$q^{ij}\partial_tK_{ij}=-\Delta\mathcal N-\mathcal NR_{\hat{N}\hat{N}}$. Keeping
$\theta_{(k)}=0$ therefore requires
\begin{equation}
-\Delta(f-\mathcal N)-|b|^2f-R_{\hat{N}\hat{N}}(f-\mathcal N)=0 .
\end{equation}
On a minimal surface in a time-symmetric vacuum slice the Gauss relation gives
$R^{(2)}=-2R_{\hat{N}\hat{N}}-|b|^2$, and $b_{AB}$ is tracefree and equal to
the shear $\sigma^{(k)}_{AB}$. Substituting $f=A+B$ and $\mathcal N=A-B$ yields
Eq.~(\ref{eq:Beqmain}). The area element changes at the rate
$(A\theta_{(k)}-B\theta_{(n)})\,\dA=-B\theta_{(n)}\,\dA$, which is
Eq.~(\ref{eq:Btransportmain}). None of these steps uses the form of the lapse.
The same steps give
$\partial_t\theta_{(n)}=2(\Delta\mathcal N+\mathcal NR_{\hat{N}\hat{N}})
=-2q^{ij}\partial_tK_{ij}=2\dot{K}_{\hat{N}\hat{N}}$, the last step because
$\gamma^{ij}\partial_tK_{ij}=-D^iD_i\mathcal N=0$ for a harmonic lapse, which is
the only place where harmonicity enters. The
coefficient of $t$ in Eq.~(\ref{eq:Btransportmain}) is therefore
$-2B\dot{K}_{\hat{N}\hat{N}}$, which is the form used numerically. For the unit
lapse $\dot{K}_{ij}=R_{ij}$, so $\dot{K}_{\hat{N}\hat{N}}=-(R^{(2)}+|\sigma^{(k)}|^2)/2$
and $f$ solves $[-\Delta+\tfrac12(R^{(2)}-|\sigma^{(k)}|^2)]f=-\dot{K}_{\hat{N}\hat{N}}$.

For the temperature lapse, write $\psi=\psi_0+O(c_1z)$ near the first hole, with
$\psi_0=a+G\mu_1/2\rho$. Then
\begin{equation}
\frac{2}{\psi_0}-1=\frac{\mathcal N_S}{a}+\mathcal N_0,\qquad
\mathcal N_S=\frac{a-G\mu_1/2\rho}{a+G\mu_1/2\rho},\quad
\mathcal N_0=\frac1a-1 .
\end{equation}
The metric $\psi_0^4\delta_{ij}$ is the Schwarzschild metric of mass
$a\mu_1=m_1$ in rescaled isotropic coordinates, and $\mathcal N_S$ is its static
lapse, which satisfies $D_iD_j\mathcal N_S=\mathcal N_SR_{ij}$ and vanishes on the
horizon. It therefore drops out of $\partial_tK_{ij}$ and of $A$ on the horizon,
and the throat evolves under the constant lapse $\mathcal N_0$, with
$B\simeq\mathcal N_0C$ and $\dot{K}_{ij}\simeq\mathcal N_0R_{ij}$; the companion's
uniform field and tide correct this at higher orders in the monopole. In the
dipole the uniform-field part of the lapse, $\mathcal N_1/\mathcal N_0=Gm_1/2\rt$
at leading order, enters at the same order as the constant-lapse term: on the
throat $R_{\hat{N}\hat{N}}=-1/R_h^2$ and $\Delta\cos\theta=-2\cos\theta/R_h^2$, so
$\partial_t\theta_{(n)}\propto-(\mathcal N_0+3\mathcal N_1\cos\theta)$. On the
round throat $R^{(2)}=2/R_h^2=1/2G^2m_1^2$, and with the monopole $C_0$ of
Eq.~(\ref{eq:Cclosedmain}) the leading monopole of the transported rate is
$\mathcal N_0^2R^{(2)}C_0\,t$, which is Eq.~(\ref{eq:mottmono}).

The equations are solved numerically on the exact minimal surface by a
Legendre--Galerkin method in fifty-digit arithmetic. For the temperature lapse the
solution of Eq.~(\ref{eq:Beqmain}) agrees with $(f-\mathcal N)/2$ to $10^{-23}$ or
better, and for the unit lapse to order $10^{-36}$. The lapse comparison of
Sec.~\ref{sec:mottmain} is identical at the resolutions $(18,64)$ and
$(24,96)$, and the Richardson extrapolations are identical at $(12,40)$ and
$(16,56)$; the first number is the Legendre truncation and the second the
number of quadrature nodes.

\section{The radial dipole of the tidal shear}
\label{app:pvdipole}

Set $G=c=1$. For a planar Newtonian orbit with separation $b$ and phase $\phi$,
let $\mathbf n$ be the unit vector from the hole to its companion,
$\mathbf e_\phi=d\mathbf n/d\phi$, and $\mathbf e_z=\mathbf n\times\mathbf e_\phi$
the normal to the orbit. The leading tidal moments at the
first hole are
\begin{equation}
\begin{gathered}
\cE_{ab}=-\frac{3m_2}{b^3}\,n_{\langle a}n_{b\rangle},\qquad
\cE_{abc}=-\frac{15m_2}{b^4}\,n_{\langle a}n_bn_{c\rangle},\\
\cB_{ab}=\pm\frac{6m_2}{b^2}\,\dot\phi\,e_{z(a}n_{b)},
\end{gathered}
\label{eq:PVmoments}
\end{equation}
where angle brackets denote the symmetric tracefree part and the sign of
$\cB_{ab}$ is the magnetic-tidal convention. Their time derivatives follow from
$\dot{\mathbf n}=\dot\phi\,\mathbf e_\phi$,
$\dot{\mathbf e}_\phi=-\dot\phi\,\mathbf n$, and
$\ddot\phi=-2\dot{b}\dot\phi/b$. Inserting them in Eq.~(\ref{eq:PVshearmain}),
with the potentials of Eq.~(\ref{eq:PVtensors}) and the odd-parity potential
$\cB^q_{AB}=2\epsilon_{apd}\,\Omega^p\cB^d{}_c\,\Omega^a_{(A}\Omega^c_{B)}$, and
contracting with the horizon metric $4m_1^2\Omega_{AB}$ gives
$\sigma_{AB}\sigma^{AB}$. Its $\ell=1$ part along $\mathbf n$ is
Eq.~(\ref{eq:generalradialmain}), the upper sign there corresponding to the
upper sign in Eq.~(\ref{eq:PVmoments}). The first group comes from the
$E_2E_3$ cross term and the second from the $E_2B_2$ cross term, which have odd
parity; the $E_2E_2$, $E_3E_3$, and $B_2B_2$ terms have even parity and no
$\ell=1$ part. The magnetic octupole, $\cB_{abc}\sim m_2V/b^4$, has no $\ell=1$
cross term with $\cE_{ab}$, by the same parity count, and enters only through
$E_3B_3$ and $B_2B_3$ cross terms, which correct the $E_2B_2$ term at relative
order $m_1^2/b^2$ and the $E_2E_3$ term at relative order $V^2$; it is omitted.
The angular average of the same expression reproduces the heating normalization
of Ref.~\cite{PoissonVlasov2010} with the magnetic octupole omitted,
$2m_1^2\langle\sigma_{AB}\sigma^{AB}\rangle=\tfrac{16}{45}m_1^6
(\dot\cE_{ab}\dot\cE^{ab}+\dot\cB_{ab}\dot\cB^{ab})
+\tfrac{16}{4725}m_1^8\dot\cE_{abc}\dot\cE^{abc}$. With the quasi-stationary
expansion $\vartheta=4m_1\sigma_{AB}\sigma^{AB}$, the $E_2B_2$ term gives
$\vartheta^{E_2B_2}_{\ell=1,{\rm rad}}=(512/5)(m_1^5m_2^2\dot{b}\dot\phi^2/b^6)
(\mathbf n\cdot\boldsymbol\Omega)$ up to sign, whence
Eq.~(\ref{eq:sixpowersmain}), and the $E_2E_3$ term at $\dot\phi=0$ gives
Eq.~(\ref{eq:headonthetamain}). We have evaluated the projection in two
independent implementations, one with three-dimensional projectors and one in
angular components with the time derivatives taken numerically along the orbit;
both reproduce Eq.~(\ref{eq:generalradialmain}) to the working precision.

\section{The strut identity}
\label{app:strut}

\begin{table}[t]
\caption{Leading-order comparison of the Brill--Lindquist and Israel--Khan
quantities of Sec.~\ref{sec:path}, under the weak-field matching
$\mu_i\leftrightarrow M_i$ (equivalently $m_i\leftrightarrow M_i$ at this order)
and $\rt\leftrightarrow\Delta z$, which is an analogy between two spacetimes and
not an equality of local constructions. The first row is
$\sum_i\bar{T}_i\,dS_i$ at fixed $\mu_i$ for Brill--Lindquist, with
$\bar{T}_i=Q_i/S_i$, and $\sum_iT_i\,dS_i$ at fixed $\lambda_i$ for Israel--Khan,
where the temperature is uniform.}
\label{tab:ikbl}
\begin{ruledtabular}
\footnotesize
\begin{tabular}{lcc}
per unit separation & Brill--Lindquist & Israel--Khan \\ \hline
first-law content & $-G\mu_1\mu_2/\rt^2$ & $-2GM_1M_2/\Delta z^2$ \\
$d(\text{total energy})$, fixed areas & $Gm_1m_2/\rt^2$ & $2GM_1M_2/\Delta z^2$ \\
mass excess (not per unit) & $Gm_1m_2/\rt$ & $2GM_1M_2/\Delta z$ \\
\end{tabular}
\end{ruledtabular}
\end{table}

Reference~\cite{CostaPerry2000} gives Eq.~(\ref{eq:IK}) for rods of lengths
$\lambda_i=2GM_i$ separated by a gap $\Delta z$, with $\kappa_1$ independent of
position on the horizon. Reference~\cite{KrtousZelnikov2020} defines the strut
length as the gap between the rod ends, which in the variables of
Eq.~(\ref{eq:IK}) is $\Delta z$, the distance between the rod centers being
$\Delta z+\tfrac12(\lambda_1+\lambda_2)$ and the half-lengths $\lambda_1/2$ and
$\lambda_2/2$. Its metric function on the strut, written there as
$(\mathsf n-2\mathsf k)/(\mathsf n+2\mathsf k)$, with $\mathsf n$ the squared
center separation less the squared half-lengths and $\mathsf k$ their product
(we have renamed their symbols to avoid collision with $\nu$ and $\sigma$ used
elsewhere here), reduces in the neutral limit to
\begin{equation}
f_{\rm s}
=\frac{\Delta z(\Delta z+\lambda_1+\lambda_2)}
{(\Delta z+\lambda_1)(\Delta z+\lambda_2)} .
\end{equation}
With $L=\Delta z\,f_{\rm s}$ and $f_{\rm s}=(1+4G\tau)^{-1}$ this gives
$\tau=\lambda_1\lambda_2/4G\Delta z(\Delta z+\lambda_1+\lambda_2)$, tending to
$GM_1M_2/\Delta z^2$ at large separation. Both sides of Eq.~(\ref{eq:strutid})
then equal Eq.~(\ref{eq:strutvalue}) with zero residual.

For the fixed-area path of Sec.~\ref{sec:path}, solving $d\A_1=d\A_2=0$ for
$d\lambda_i/d\Delta z$ and substituting gives
$d\mathcal H/d\Delta z=\lambda_1\lambda_2/2G\Delta z^2+\cdots$, while
$-L\,d\tau/d\Delta z$ evaluated along the same path gives the same expression, so
the first law closes with zero residual there also. Along that path
$\sum_iT_i\,dS_i=0$ identically, the areas being held fixed.

One comparison with Sec.~\ref{sec:rigid} is worth recording. Solving
$\A_i=16\pi G^2m_i^2$ for an irreducible mass in each solution and asking by how
much the sum exceeds the total energy, Eq.~(\ref{eq:misner}) gives exactly one
Newtonian binding energy, $E_b$, in Brill--Lindquist, whereas Israel--Khan gives
twice that at leading order; the additional $GM_1M_2/\Delta z$ equals $\tau L$ at
that order. At leading order $-\tau L$ is the redshifted energy of a conical
excess, which is the same fact that makes $L$ the thermodynamic length; we do not
press the reading further than that. We attach no significance to the comparison
beyond leading order, since $\rt$ and $\Delta z$ are coordinate separations in
different spacetimes and the order at which the agreement ends depends on which
masses are identified with one another. Table~\ref{tab:ikbl} collects the
leading-order comparison used in Sec.~\ref{sec:path}.

\end{document}